\documentclass[article,twocolumn,preprintnumbers,superscriptaddress,amsmath,amssymb,aps,longbibliography,nofootinbib]{revtex4-2}
\RequirePackage[colorlinks=true,urlcolor=blue,anchorcolor=blue,
  citecolor=blue,filecolor=blue,linkcolor=blue,menucolor=blue,
  linktocpage=true,pdfproducer=medialab,pagebackref=true]{hyperref}
\hypersetup{colorlinks=true,urlcolor=blue,anchorcolor=blue,
  citecolor=blue,filecolor=blue,linkcolor=blue,menucolor=blue,
  pdfproducer=medialab}
\usepackage[utf8]{inputenc}
\usepackage[table,dvipsnames]{xcolor}
\usepackage{colortbl}
\usepackage{ragged2e}
\usepackage[english]{babel}
\usepackage{bbold}
\usepackage[noindentafter]{titlesec}
\usepackage{booktabs}
\usepackage{physics}
\usepackage{booktabs}
\usepackage{braket}
\usepackage{slashed}
\usepackage{orcidlink}
\usepackage{indentfirst}
\usepackage{graphicx}
\usepackage{float}
\usepackage{enumerate}
\usepackage{subcaption}
\usepackage{color}
\definecolor{red}{rgb}{1,.0706,.1373}
\definecolor{blue}{rgb}{0,0.396,0.741}
\definecolor{realblue}{rgb}{0,0,1}
\definecolor{green}{rgb}{0.25,0.6,0.2}
\definecolor{rossoc}{cmyk}{0,1,1,0.2}
\definecolor{dpink}{cmyk}{0,0.58,0.55,0.14}

\definecolor{teal}{rgb}{0, 0,0.8}
\colorlet{mylinkcolor}{teal}
\colorlet{mycitecolor}{teal}
\colorlet{myurlcolor}{teal}
\hypersetup{
  linkcolor  = mylinkcolor!,
  citecolor  = mycitecolor!,
  urlcolor   = myurlcolor!,
  colorlinks = true
}

\newcommand{\hc}{\; + \; \mathrm{h.c.} \;}

\begin{document}
\title{
The Dark Dimension and Majorana Neutrinos
}
\preprint{IPPP/26/63}
\author{Arturo de Giorgi~\orcidlink{0000-0002-9260-5466}
}
\email{arturo.de-giorgi@durham.ac.uk}
\author{Dhruv Pasari~\orcidlink{0009-0007-1283-1492}
}
\email{dhruv.pasari@durham.ac.uk}
\affiliation{Institute for Particle Physics Phenomenology, Department of Physics, Durham University, Durham DH1 3LE, U.K.}
\begin{abstract}
Recent developments in the Swampland program motivate the existence of a mesoscopic Dark Dimension of size $0.1-10~\mu$m with bulk Majorana fermions in the $1-10$~keV range. Motivated by this, we derive terrestrial constraints on Majorana bulk neutrinos as a function of the compactification radius and confront them with the preferred region of the Dark Dimension scenario. After determining the mass spectrum and mixing structure, we confront the model with Daya Bay neutrino-oscillation data, the KATRIN beta-decay bound, and the KamLAND-Zen neutrinoless double beta decay limit. For the latter, we obtain a closed-form expression for the effective Majorana mass and show that the Kaluza--Klein tower efficiently screens neutrinoless double beta decay when the nuclear momentum exceeds the compactification and Majorana scales. In the Dark-Dimension window, beta decay provides the strongest constraint, and, together with the measured neutrino mass splittings, it restricts the largest brane Yukawa coupling to $\mathcal{O}(10^{-3})$, with a weak dependence on the compactification scale. Oscillation and neutrinoless double beta decay searches remain complementary, probing smaller and larger Majorana masses, respectively. For completeness, outside the Dark-Dimension assumptions, we investigate how hierarchical Majorana masses can weaken the constraints.
\end{abstract}
\maketitle
\tableofcontents
\section{Introduction}

In the past few years, it was argued that Swampland-motivated arguments about the consistency of the low-energy effective theory with quantum gravity, together with the observed value of the cosmological constant, single out a scenario where the Standard Model~(SM) is confined to a 4D brane, alongside a mesoscopic extra Dark Dimension~(DD) of order a micron~\cite{Montero:2022prj}
\begin{align}
    \label{eq:DD-pred}  &R_\text{DD}\sim 0.1-10~\mu\text{m}\,,    &R_\text{DD}^{-1}\sim 20~\text{meV}-2~\text{eV}\,,
\end{align}
alongside a tower of bulk fermions associated with it. The striking numerical coincidence of Eq.~\eqref{eq:DD-pred} with the scale of the active neutrino masses suggestively hints at the identification of such bulk states with the natural candidate for right-handed neutrinos.

The most direct experimental signature of a DD is a modification of Newton’s law of gravitation. The strongest constraints on such modifications come from torsion-balance and Casimir-force experiments at submillimeter distances~\cite{Hoskins:1985tn,Bordag:2001qi,Mostepanenko:2001fx,Chiaverini:2002cb,Long:2003dx,Chen:2014oda,Tan:2016vwu,Tan:2020vpf}~(see Ref.~\cite{Murata:2026llo} for a review). For a flat extra dimension, the leading bound up to date comes from the torsion-balance Washington~2020 measurement~\cite{Lee:2020zjt}
\begin{align}
    \label{eq:fifth-forces}&R<30~\mu\text{m}\,, &R^{-1}>6.6~\text{meV}\,.
\end{align}
The constraint is remarkably close to the DD preferred area, yet it does not suffice to cover it.

On the other hand, neutrinos can provide a remarkably sensitive probe of extra dimensions, in some regimes offering greater sensitivity than direct gravitational tests. The bulk sterile fermions can mix with the SM active neutrino on the brane, thus participating in neutrino mass generation as effective ``right-handed neutrinos''~\cite{Dienes:1998sb,Arkani-Hamed:1998wuz,Dvali:1999cn,Lukas:2000rg}. Therefore, one may naturally expect neutrino physics to be sensitive to extra dimensions with a compactification radius comparable to the neutrino mass scale, $R^{-1}\sim m_\nu$. Indeed, this turns out to be the case. For example, oscillation experiments can probe compactification scales well below the reach of fifth-force experiments down to $R\lesssim 0.6~\mu\text{m}$ for Normal Ordering~(NO) and $R\lesssim 0.1~\mu\text{m}$ for Inverted Ordering~(IO) at $99\%$~C.L.~\cite{Forero:2022skg,Elacmaz:2025ihm,deGiorgi:2025xgp,Franklin:2025muw}, about $50-300$ times stronger!
Further experiments and motivated extensions of the DD have been considered beyond the vanilla scenario, leading to a rich literature about the topic~\cite{Antoniadis:2025rck, Eller:2025lsh, Bai:2026kdq,Langhoff:2026qqg}.

Despite the proposal's appealing features, which elegantly link the hierarchy of scales, some of them might appear ad hoc, as recognised by the same proposers~\cite{Montero:2022prj,Montero:2025hye}. Giving mass to the three active neutrinos requires postulating at least three massless bulk fermions, with nothing in the construction fixing this number beyond the phenomenological need to match the observed neutrino flavour structure. Moreover, the framework leaves $B-L$ global symmetry of the SM merely accidental, without any mechanism protecting it, in tension with the expectation that quantum gravity does not tolerate exact global symmetries. Motivated by this, it was proposed to extend the DD by promoting $B-L$ to a bulk gauge symmetry~\cite{Montero:2025hye}. There are two benefits: i) $B-L$ becomes a genuine, protected symmetry of the SM, and ii) three generations of bulk right-handed neutrinos are \textit{predicted}, rather than postulated, fixed by anomaly inflow cancellation. Within the setup, the model predicts a set of Majorana masses for the bulk fermions in the keV range~\footnote{
For the sake of completeness, we point out that the Majorana neutrinos and their mass range presented in Ref.~\cite{Montero:2025hye} come with further particle content and some limitations. First, the theory contains a bulk scalar charged under $B$-$L$ and the associated massive $B$-$L$ gauge boson benchmarked by the same naturalness arguments at $m_{B\text{-}L}\sim100$~GeV and $g_{B\text{-}L}\sim10^{-10}$. We note, however, that $U(1)_{B\text{-}L}$ is not orthogonal to $U(1)_Y$ in the SM, since the SM fermions are charged under both. Therefore, a kinetic-mixing term $\epsilon\,F^{B\text{-}L}_{\mu\nu}F^{Y\,\mu\nu}$ is generically induced. Its value depends on the energy scale at which $\epsilon$ is assumed to vanish and on the renormalization-group running down from there, so the resulting magnitude is not fixed by the model. A similar discussion applies to the mass ranges: $M$ and $m_{B\text{-}L}$ both arise from the same $B$-$L$-breaking VEV, but through independent Yukawa and gauge couplings respectively, so nothing in the model fixes their relative size either, and the predicted keV range for $M$ rests on naturalness assumptions about such couplings. Therefore, the quoted mass ranges have to be understood as phenomenological benchmarks. All in all, the full model is subject to phenomenological constraints on the mixing and on the $B$-$L$ gauge boson beyond those considered in Ref.~\cite{Montero:2025hye} or in this work. A compilation of constraints on both the kinetic mixing and on the $g_{B-L}$ can be found, e.g. in Ref.~\cite{AxionLimits}.}

\begin{equation}
    M_\text{DD}\sim 1-10~\text{keV}\,.
\end{equation}
Such a value of the Majorana mass comfortably sits in the testable region by terrestrial neutrino experiments and motivates us to study this scenario, in a quest to shed light on the Dark Dimension.

Somewhat surprisingly, a bulk Majorana neutrino subject to a 5D-Lorentz-invariant (5DLI) mass condition has not, to our knowledge, received a dedicated in-depth phenomenological treatment. We believe the origin of this has to be traced to the model-building intent of the authors when choosing the definition of charge conjugation. The ordinary 4D charge-conjugation matrix $C$ does not by itself furnish a 5DLI Majorana-type mass term for a bulk fermion $\bar{\Psi}\Psi^c$, where $\Psi^c\equiv C\bar{\Psi}^T$ is the charge-conjugated spinor. The 5DLI bilinear is instead built from $\Psi^{5c}$ defined via the operator  $C_5=\gamma^5 C$. This point was discussed in the context of allowed 5DLI operators in Ref.~\cite{Lukas:2000rg,Pilaftsis:1999jk} and stated explicitly more recently in Ref.~\cite{Abe:2026jtf}. Unlike the ordinary 4D charge-conjugation matrix, for which $(\Psi^c)^c=\Psi$, the double conjugation of a bulk spinor closes onto $(\Psi^{5c})^{5c}=-\Psi$ rather than $+\Psi$ (cf.\ App.~\ref{app:model}), thus undermining the usual interpretation which gives the name ``charge conjugation''. 
The correct 5DLI bilinear must be built and interpreted accordingly. We believe this subtlety has led much of the subsequent phenomenological work to the ordinary 4D operator $C$~\cite{Dienes:1998sb,Bhattacharyya:2002vf,deGiorgi:2025xgp,Blennow:2010zu} instead of $C_5$.
\bigskip

In this work, we fill this gap. We derive the 4D KK masses and mixings from the genuinely 5DLI bulk Majorana mass term, and derive the corresponding phenomenology. We study three of the most relevant laboratory constraints: flavour oscillation, neutrinoless double beta decay ($0\nu\beta\beta$), and beta decay\footnote{For a similar analysis on a different model of Naturalness with $N$ sterile neutrinos, see e.g. Ref.~\cite{Lonardi:2026exa}.}. This allows us to set new stringent limits on the bulk Majorana mass and the compactification scale $R$ as a function of the Dirac Yukawa couplings.

We find that these observables provide complementary probes of different regions of parameter space. In particular, somewhat counter-intuitively, the KK tower effectively screens $0\nu\beta\beta$ when many states lie below the characteristic nuclear momentum scale. The expressions and resulting bounds are derived without including any Swampland conjecture, and hence our results retain validity beyond the DD scenario. Within the DD-favoured parameter space, beta decay provides the strongest constraint.  
We find that viable DD scenarios require brane Yukawa couplings of order $10^{-3}$, implying a mild hierarchy with only a weak scaling of $R^{-1/6}$.
It is worth noting that the DD proposal relies heavily on naturalness arguments. As we discuss later, large hierarchies among the model's parameters can alter the available parameter space.

\bigskip
The paper is structured as follows. In Sec.~\ref{sec:model}, we present the model and derive the particle spectrum and mixings relevant for phenomenology. In Sec.~\ref{sec:observables}, we then list the relevant observables, experiments, and expressions which will be employed to set constraints. The details about the analysis can be found in Sec.~\ref{sec:analysis}, and in Sec.~\ref{sec:results} we present the results. Finally, conclusions are drawn in Sec.~\ref{sec:conclusions}. Further supplemental material can be found in the Appendices.
\section{The 5D Model and the 4D Spectrum}
\label{sec:model}
\subsection{The Model}
The model we study introduces a circular orbifolded $S_1/\mathbf{Z}^2$ flat extra dimension, $y$, of radius $R$~\cite{Antoniadis:1990ew,Arkani-Hamed:1998jmv,Antoniadis:1998ig}.
The fundamental cutoff of the theory is dictated by the five-dimensional Planck mass $M_*$. This is related to the 4D one by the volume of the extra dimension $V=2\pi R$ via $\bar{M}_{\rm Pl}^2=V\,M_*^3$, where $\bar{M}_{\rm Pl}\approx 2.4\times 10^{18}$~GeV is the reduced Planck mass.
We consider the case in which the SM cannot propagate in the extra dimension, and it is confined on a brane at one of the orbifold fixed points $y=0,\pi R$; for the sake of concreteness, we choose $y=0$.
Within the extra dimension, only gravity and three Majorana bulk fermions $\Psi_i$ can propagate. Their free action is given by
\begin{equation}
S_{\rm bulk} = \int d^4x\int_{-\pi R}^{\pi R} dy\,\left[
i\overline\Psi\Gamma^A\partial_A\Psi - \frac{1}{2}\overline\Psi \mathbf{M}\Psi^{5c}\hc\right]\,,
\label{eq:S5c}
\end{equation}
where $\mathbf{M}$ is the Majorana mass matrix, and we omitted flavour indices.
The fields are conventionally assigned positive orbifold parity $\gamma^5 \Psi_i(x,-y)=+\Psi_i(x,y)$, which induces a parity on the Weyl component of the spinor $\Psi=\Psi_L+\Psi_R$
\begin{align}
   \label{eq:parity}&\Psi_L(x,-y)=-\Psi_L(x,y)\,, &\Psi_R(x,-y)=+\Psi_R(x,y)\,.
\end{align}
More details about the conventions adopted in this work can be found in App.~\ref{app:model}.
The 5D charge conjugated spinor $\Psi^{5c}$ is defined as
\begin{align}
&\Psi^{5c} \equiv C_5\overline\Psi^T
&&C_5 \equiv \gamma^5 C_4\,, &&C_4=i\gamma^2\gamma^0\,,
\label{eq:C5def}
\end{align}
where $C_4$ is the usual 4D charge conjugation.

The bulk fermion couples to the SM via the Yukawa neutrino portal on the brane. Given the orbifold parity assignment, $\Psi_L(y=0)=0$, and only the right-handed component can couple to the SM. Above electroweak symmetry breaking~(EWSB), the relevant 5D operator localised at the brane is given by
\begin{equation}
    S_{\rm br}\supset -\int d^4x\,dy\,\delta(y)\,\Bar{L}\Tilde{H}\frac{\mathbf{y}_D}{\sqrt{M_*}}\Psi_R(x,y)+{\rm h.c.}\,,
\end{equation}
where $L$ is the leptonic EW doublet, $\tilde H$ is the Higgs dual, and $\mathbf{y}_D$ is the dimensionless 5D Dirac Yukawa coupling.
In the following, we take the simplified path of assuming that $\mathbf{M}$ and $\mathbf{y}_D$ are simultaneously diagonalisable. While this is unlikely to be the case in a realistic model without a flavour symmetry, it ought to capture the size of the constraints.

\subsection{The 4D Spectrum}
The 4D EFT can be obtained by integrating out the extra dimension. We report the main findings below, while details about the derivation can be found in App.~\ref{app:model-spectrum}. After EWSB, $\langle H^\dagger H\rangle=v^2/2$, with the Higgs vev $v\approx 246$~GeV. Upon performing the KK expansion of the modes, it yields the brane-operator
\begin{equation}
    S_{\rm br}\supset -\int d^4x\,\bar{\nu}_{L}\mathbf{m}_D\left(\Psi_{R,0}(x)+\sum\limits_{n=1}^\infty \sqrt{2}\Psi_{R,n}\right)+{\rm h.c.}\,,
    \label{eq:action-mixing}
\end{equation}
where we defined the 4D Dirac mass matrix
\begin{equation}
    \mathbf{m}_D\equiv \frac{v}{\sqrt{2}}\frac{\mathbf{y}_D}{\sqrt{VM_*}}= \frac{\mathbf{y}_Dv}{\sqrt{2}}\times\left(\frac{M_*}{\bar{M}_{\rm Pl}}\right)\,.
    \label{eq:mD-yukawa-DD}
\end{equation}
In terms of naturalness, in the DD one expects Dirac masses of the order
\begin{equation}
 m_D
 \approx(70~\mathrm{eV})
 \times y_D~\left(\frac{1/R}{\mathrm{eV}}\right)^{1/3}\,,
\label{eq:mD-yukawa-DD-natural}
\end{equation}
which depends very mildly on the size of the extra dimension.
Allowing $y_D$ to span down to a reasonable hierarchy of $\mathcal{O}(10^{-2})$, this corresponds approximately to the range $m_D\in [0.1,100]$~eV.

\bigskip
The operator defined in Eq.~\eqref{eq:action-mixing} mixes the SM neutrino with the whole KK tower. We choose to work in the basis in which both $\mathbf{m}_D$ and $\mathbf{M}$ are diagonal. This reduces the problem from three flavours to three copies of the single-flavour case, then linked by the PMNS matrix, allowing us to omit flavour indices for the time being. We can encode all states within a vector $X$, such that the mixing of the KK states with the SM neutrino is encoded in the mass matrix $-\mathcal L\supset\tfrac12\overline X\,\mathcal M\,X^c+\mathrm{h.c.}$
with
\begin{widetext}
\begin{align}
&\mathcal{M} =
\left(\begin{array}{ccccccc}
0 & m_D & \sqrt2m_Ds_1 & \sqrt2m_Dc_1 & \dots & \sqrt2m_Ds_N & \sqrt2m_Dc_N \\
m_D & -M & 0 & 0 & \dots & 0 & 0 \\
\sqrt2m_Ds_1 & 0 & D_1 & 0 & \dots & 0 & 0 \\
\sqrt2m_Dc_1 & 0 & 0 & -D_1 & \dots & 0 & 0 \\
\vdots & \vdots & \vdots & \vdots & \ddots & \vdots & \vdots \\
\sqrt2m_Ds_N & 0 & 0 & 0 & \dots & D_N & 0 \\
\sqrt2m_Dc_N & 0 & 0 & 0 & \dots & 0 & -D_N
\end{array}\right)\,,
    & \begin{matrix}
    s_n^2=\frac{D_n-M}{2D_n}\,,\\\\ c_n^2=\frac{D_n+M}{2D_n}\,,\\
    \\
    D_n\equiv\sqrt{M^2+\mu_n^2}\,,
\end{matrix}
\label{eq:massmatrix_final}
\end{align}
\end{widetext}
where $\mu_n\equiv n/R$ and $N\gg 1$ is an arbitrary truncation point. In practice, we will take $N\to\infty$ to obtain analytical results, while for numerical estimations one needs to ensure $N$ is sufficiently large not to introduce unphysical features in the part of the spectrum relevant for the observables.

As can be seen from the mass matrix of Eq.~\eqref{eq:massmatrix_final}, the contribution of the extra-dimensional states beyond the mixing is given by the mass terms $D_n=\sqrt{M^2+\mu_n^2}$, matching previous results in the literature~\cite{Lukas:2000rg,Kim:2003vr,Eisele:2006va,Watanabe:2010cy,Abe:2026jtf,Diego:2008zu}. This is nothing but the contribution to the usual relativistic energy-mass relations stemming from the 5DLI theory, $E^2=\vec{p}_4^2+\vec{p}_5^2+M^2$, with $\vec{p}_{4,5}$ the three- and fifth-dimensional momenta, where $p_5$ gets quantised in units of $1/R$ due to the orbifold.  

The physical masses, $\lambda_n$, correspond to the eigenvalues of the mass matrix $\mathcal{M}$. They can be conveniently computed as roots of the secular function (cf.~App.~\ref{app:eigensystem})
\begin{align}
\label{eq:eigenvalue-eq}&f(\lambda)\equiv \lambda - \frac{\pi m_D^2(\lambda-M)}{\mu_1\,w}\,\cot\!\left(\frac{\pi w}{\mu_1}\right)\,,
\end{align}
where $w=\sqrt{\lambda^2-M^2}$. Notice that $w$ is imaginary when $\lambda<M$, as for the lightest mode.
The mixing of the mass eigenstate $\lambda$ with the SM neutrino $\mathcal{N}_\lambda$ can be computed in terms of the derivative of $f(\lambda)$ yielding $\mathcal N_\lambda^{-2}=f'(\lambda)$ (cf.~App.~\ref{app:eigensystem}) and thus
\begin{equation}
    \mathcal N_\lambda^{-2} = \frac{\lambda^2-M\lambda-M^2}{w^2}
    +\frac{\pi^2m_D^2\,\lambda(\lambda-M)}{\mu_1^2\,w^2}\,\csc^2\!\left(\frac{\pi w}{\mu_1}\right)\,.
\end{equation}
Furthermore, due to unitarity, the mixings satisfy, for each flavour, the sum rule
\begin{equation}
    \sum\limits_\lambda\mathcal{N}_\lambda^2=1\,.
\end{equation}
Such a relation can be used to quantify whether the truncation at $N$ is acceptable for reliably computing observables.
The masses ${\lambda_n}$ and mixings $\mathcal{N}_\lambda$ so derived can then be employed to predict all the relevant observables.
\section{Observables}
\label{sec:observables}
In this work, we consider three laboratory observables: neutrino flavour oscillations, neutrinoless double beta decay, and tritium beta decay. All of them are computed in the three-flavour scenario.
For this, before proceeding, we clarify the notation and technical aspects related to the flavour indices neglected until now.
Throughout, $\alpha\in\{e,\mu,\tau\}$ denotes flavour, $j,k,\dots\in\{1,2,3\}$ the generation (tower) index, and $n,m,\dots\in\{0,1,\dots,N\}$ the KK level, with $n=0$ the brane zero mode.

In the full three-flavour picture, each lepton doublet $\alpha$ couples to the bulk field of every generation through the brane operator, with Yukawa matrix $Y_{\alpha j}\propto U_{\alpha j}\,m_{D,j}$, where $U_{\alpha j}$ is the (PMNS) matrix rotating the three flavours into the three towers and simultaneously diagonalising $Y_D$, $m_{D,j}$ is the Dirac mass of tower $j$, and each tower has its own bulk Majorana mass $M_j$.

In the basis in which $Y_D$ has the singular-value form $Y_D=U\,\mathrm{diag}(m_{D,1},m_{D,2},m_{D,3})$, the full mass matrix is simply the direct sum of three decoupled copies of the single-tower matrix already studied,
\begin{equation}
\tilde{\mathbf M}=\bigoplus_{j=1}^3\mathbf M^{(j)}\,,
\end{equation}
with no coupling between different generations beyond the PMNS matrix. All of the flavour mixing has been absorbed into a unitary rotation $O=\mathrm{diag}(U,\mathbb 1)$, which acts only on the three flavour rows/columns and leaves every KK row untouched, so that $\mathbf M=O\,\tilde{\mathbf M}\,O^T$.
As discussed in App.~\ref{app:mixing-SM}, the matrices $\mathbf M^{(j)}$ are still non-diagonal; reaching the actual propagating mass eigenstates $\hat\nu_{j,m}$ of tower $j$ requires a second, generation-dependent rotation $V^j_{nm}$, generically different from tower to tower since each has its own $m_{D,j},M_j$. In terms of the interaction (flavour) states $\nu_{\alpha,n}$, with $n=0$ the brane zero mode, i.e. the physical SM doublet $\nu_\alpha\equiv\nu_{\alpha,0}$, the full change of basis then reads
\begin{equation}
\label{eq:basis-definition}
\ket{\nu_{\alpha,n}}=\sum_j\sum_m U_{\alpha j}\,V^j_{nm}\,\ket{\hat\nu_{j,m}}\,,
\end{equation}
with $U_{\alpha j}$ carrying no KK index and all the KK-level dependence, together with its generation dependence, sitting in $V^j_{nm}$. The exact form of $V^j_{nm}$ is determined by the eigenvalues of the mass matrix, and can be found in App.~\ref{app:eigensystem} in Eq.~\eqref{eq-app:eigenvector_closed}, tower by tower.

\bigskip
Before proceeding with the observables, it is useful to notice a simple property of the spectrum. The eigenvalue equation of Eq.~\eqref{eq:eigenvalue-eq} can be rearranged as
\begin{equation}
m_D^2 =
\frac{\mu_1\lambda w}{\pi(\lambda-M)}
\tan\!\left(\frac{\pi w}{\mu_1}\right),
\qquad
w=\sqrt{\lambda^2-M^2},
\label{eq:mD-inverted}
\end{equation}
The lightest eigenvalue $\lambda_0$ is continuously connected to zero at $m_D=0$ and approaches the first zero of the cotangent, $w=\mu_1/2$, as $m_D\to\infty$. Hence, its exact range is
\begin{equation}
0\leq\lambda_0<
\sqrt{M^2+\left(\frac{\mu_1}{2}\right)^2}.
\label{eq:light-root-range}
\end{equation}
For universal $M$, all three generations share the same
upper bound. Reproducing the largest measured splitting therefore requires
\begin{equation}
m_{\rm light}^2+\Delta m_{\max}^2<{M^2+\left(\frac{\mu_1}{2}\right)^2},
\end{equation}
where $\Delta m_{\max}^2=\Delta m_{31}^2$ for normal ordering and
$\Delta m_{\max}^2=|\Delta m_{32}^2|$ for inverted ordering.
Since $m_{\rm light}=0$ is allowed, a necessary condition
for spectral reachability is
\begin{equation}
M^2+\left(\frac{\mu_1}{2}\right)^2>\Delta m_{\max}^2.
\label{eq:spectrum-necessary}
\end{equation}
This qualitative result, indeed shows the power of neutrino physics in constraining the extra-dimensional parameters: regions violating this inequality cannot reproduce the measured mass splittings for any finite choice of the Dirac mixings.

\subsection{Oscillations}
\begin{figure}[t]
    \centering
    \includegraphics[width=0.49\textwidth]{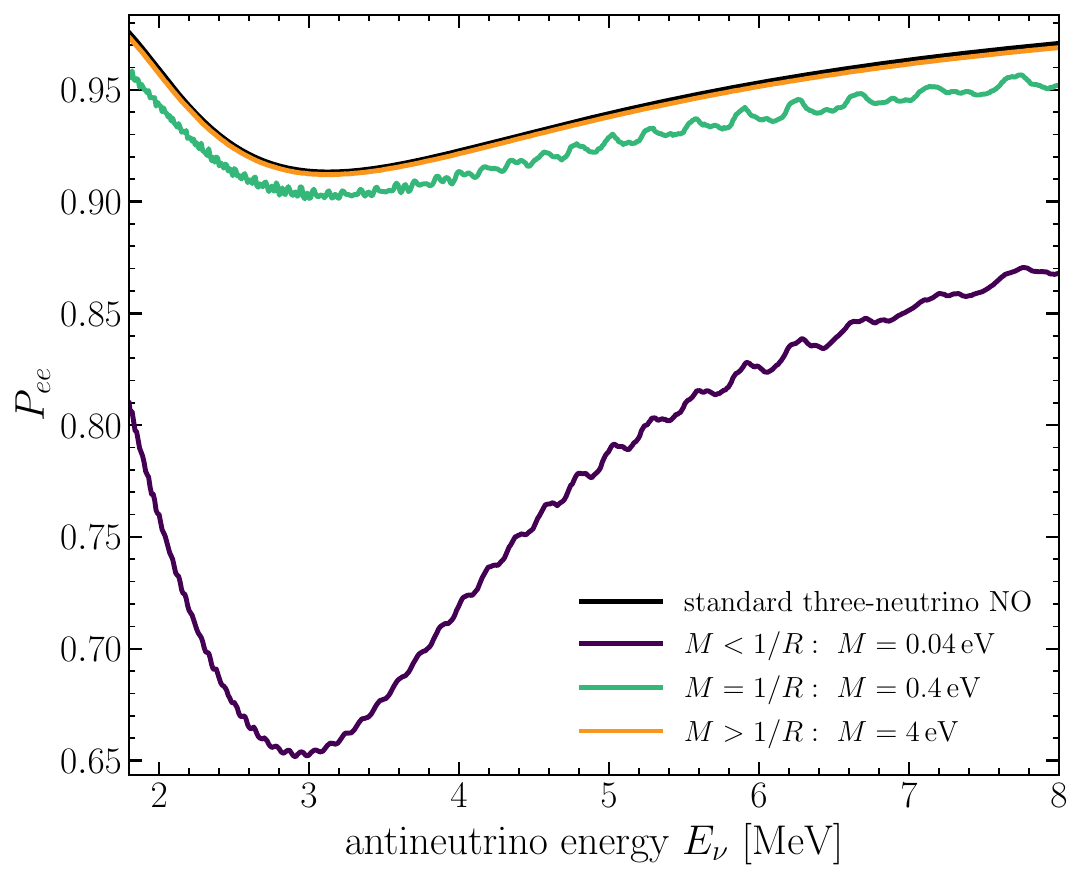}
    \caption{Baseline-averaged electron-antineutrino survival probability in
    NO. The standard three-neutrino prediction is compared with three exact
    KK benchmarks at fixed $\mu_1=1/R=0.4~\mathrm{eV}$ and
    $m_{D,1}=0.01~\mathrm{eV}$, with universal Majorana masses
    $M=0.04$, $0.4$, and $4~\mathrm{eV}$. For every benchmark, the remaining
    Dirac masses are fixed to reproduce the measured NO mass splittings. The
    probability is averaged with inverse-square weights over the 18
    reactor--detector baselines of the Daya Bay six-AD far hall.}
    \label{fig:oscillation-benchmarks}
\end{figure}
\paragraph{Observable.}
The first observable we consider is neutrino flavour oscillation. We consider the Daya Bay experiment, whose short baselines and high-statistics reactor sample provide one of the most stringent bounds on extra-dimensional neutrino oscillation phenomenology~\cite{Forero:2022skg,Elacmaz:2025ihm,deGiorgi:2025xgp} \footnote{We choose not to include MINOS/MINOS+~\cite{MINOS:2017cae}. A consistent long-baseline treatment of this model requires propagation in matter, whereas the resulting correction is relevant mainly in the low-$M$ region and does not materially affect the keV Majorana window central to the DD scenario. }. The complete 3158-day data set contains $5.55\times10^6$ inverse-beta-decay candidates collected with detectors at baselines of approximately $0.36$-$2.0$~km~\cite{DayaBay:2022orm,DayaBay:2024nip}. Daya Bay can be treated in vacuum to excellent accuracy and therefore provides the clean oscillation input used in this work without the need for matter effects.

The oscillation probability in vacuum $P_{\alpha\to\beta}(t) \equiv|\braket{\nu_{\beta,0}(t)|\nu_{\alpha,0}(0)}|^2$ of a SM flavour eigenstate $\ket{\nu_{\alpha,0}}$ in time is given by
\begin{align}
    &P_{\alpha\to\beta}(t) =\left|\sum\limits_{i,n}\left(U_{\alpha i}V^{i}_{0n}\right)\left(U_{\beta i}V^{i}_{0n}\right)^\star e^{-i E_{in}t}\right|^2 \,,
\end{align}
where $E_{in}\simeq E+\frac{m_{in}^2}{2E}$ is the energy of the mass eigenstate $\ket{\hat\nu_{i,n}}$ in the ultra-relativistic limit. 

The impact of the KK tower on the electron-antineutrino survival probability is illustrated in Fig.~\ref{fig:oscillation-benchmarks}. We fix $\mu_1=1/R=0.4~\mathrm{eV}$ and $m_{D,1}=0.01~\mathrm{eV}$, and choose three representative values of the universal Majorana mass spanning $M<\mu_1$, $M=\mu_1$, and $M>\mu_1$. For each benchmark, the remaining Dirac masses are fixed by the measured NO mass splittings. The departure from the standard three-neutrino prediction is largest for $M<\mu_1$, where a sizeable fraction of the active state is distributed over the KK tower. As $M/\mu_1$ increases, the light mode saturates the active norm and the standard oscillation pattern is recovered. The small rapid oscillations superimposed on the usual atmospheric-scale shape are generated by the additional KK frequencies and are partially washed out by the average over the far-hall baselines.

\bigskip
\paragraph{Qualitative Constraint.}
In the DD scenario, the regime of interest is $M\gg\mu_1,m_D$. In this limit, many states become weakly coupled, but almost degenerate in mass, potentially enhancing the signal. Solving for the lightest eigenvalue via Eq.~\eqref{eq:eigenvalue-eq} gives
\begin{equation}
    \label{eq:lightest}\lambda_{i,0} \approx \frac{\pi m_{i,D}^{2}}{\mu_1}\,, \qquad
    \left(V^{i}_{00}\right)^2 \approx \left(1+\frac{\pi m_{i,D}^{2}}{\mu_1 M_i}\right)^{-1}\,.
\end{equation}
Since $\left(V^i_{00}\right)^2$ remains close to unity throughout this regime, the lightest mode dominates the completeness sum, $\sum_n\left(V^i_{0n}\right)^4\approx\left(V^i_{00}\right)^4$, and the requirement of near-unitary oscillations, $\sum_n\left(V^i_{0n}\right)^4\gtrsim1-\epsilon$, translates into the naive bound
\begin{equation}
     \mu_1\gtrsim \frac{2\pi m_D^2}{\epsilon M}\propto \frac{1}{M}\,.
     \label{eq:scaling-oscillations}
\end{equation}
Here $\epsilon$ is related to the precision of the experiment, and it is therefore expected to be $\mathcal{O}(1\%)$.
All in all, one expects the bound on $R$ to become weaker as $M$ gets larger.

\subsection{Neutrinoless Double Beta Decay}
\label{sec:0nubb}

\paragraph{Observable.}
The characteristic prediction of a Majorana mass in the theory is the existence of lepton-number-violating processes, one of the most studied being neutrinoless double beta decay $0\nu\beta\beta$ (see Refs.~\cite{Rodejohann:2011mu,Dolinski:2019nrj} for a review).
The current most stringent constraint comes from the complete KamLAND-Zen data, which constrain the half-life of the process~\cite{KamLAND-Zen:2024eml}
\begin{equation}
 T_{1/2}^{0\nu}(^{136}{\rm Xe})>3.8\times10^{26}~{\rm yr}
 \qquad (90\%~{\rm C.L.}) .
\end{equation}
To translate it to a bound on $m_{\beta\beta}$, we employ the isotope-matched interpolation~\cite{Blennow:2010th,Abada:2018qok}
\begin{align}
&m_{\beta\beta}\approx \left|\sum_k \mathbf U_{ek}^2m_k
\frac{p^2}{p^2+m_k^2}\right|\,,
\label{eq:mbb-interpolation}
\end{align}
where $p\equiv\sqrt{\langle p^2\rangle}\sim \mathcal{O}(100)$~MeV is the associated nuclear momentum scale.
Possible Majorana phases are encoded in both the masses and the mixings and may lead to cancellations.
The nuclear response is an important source of uncertainty for this process. For two internally matched, unquenched $^{136}$Xe QRPA calculations, we consider the Argonne and CD--Bonn results~\cite{Faessler:2014kka}. The corresponding values of $p$ and the subsequent value of $m_{\beta\beta}^{\rm lim}$ are reported in Tab.~\ref{tab:mbb-values}.
\begin{table}[h]
\centering
\begin{tabular}{l|ccccc}
\toprule
 Method &$\left.\qquad\right.$& $p~[\mathrm{MeV}]$ &$\left.\qquad\right.$& $m_{\beta\beta}^{\rm lim}~[\mathrm{meV}]$&$\left.\qquad\right.$ \\
\midrule
Argonne  && 183.0 && 62.0&\\
CD--Bonn && 210.8 && 54.9& \\
\bottomrule
\end{tabular}
\caption{Values of $p$ and the derived $m_{\beta\beta}^{\rm lim}$ for two internally matched, unquenched $^{136}$Xe QRPA calculations.}
\label{tab:mbb-values}
\end{table}
The explicit numerical conversion, including the matched light- and heavy-exchange matrix elements, is given in App.~\ref{app:xe136-conversion}.

In the presence of the KK towers, the definition of Eq.~\eqref{eq:mbb-interpolation} is extended to
\begin{equation}
m_{\beta\beta}\approx \left|\sum_{j}\sum_m \big(U_{e j}V^j_{0m}\big)^2\,\lambda_{j,m}\,\frac{p^2}{ p^2+\lambda_{j,m}^2}\right|\,.
\label{eq:mbb_spectral}
\end{equation}
The sum over the KK states, both light and heavy, can be computed exactly, and we obtain the master formula
\begin{equation}
m_{\beta\beta}=\left|\sum_{j=1}^3U_{ej}^2\,\mathrm{Re}\left[\frac{p^2}{f^{(j)}(ip)}\right]\right|\equiv \left|\sum_{j=1}^3U_{ej}^2\,m_{\beta\beta,j}(p)\right|\,,
\label{eq:mbb_general}
\end{equation}
where $f^{(j)}(z)$ is the same secular function already defined in Eq.~\eqref{eq:eigenvalue-eq} for each flavour $j$. A detailed derivation of the formulas can be found in App.~\ref{app:0vbb}.
The quantity constrained by $0\nu\beta\beta$ is the coherent three-tower
combination $m_{\beta\beta}=|\sum_j U_{ej}^2m_{\beta\beta,j}|$.

\begin{figure}[t]
    \centering
    \includegraphics[width=0.49\textwidth]{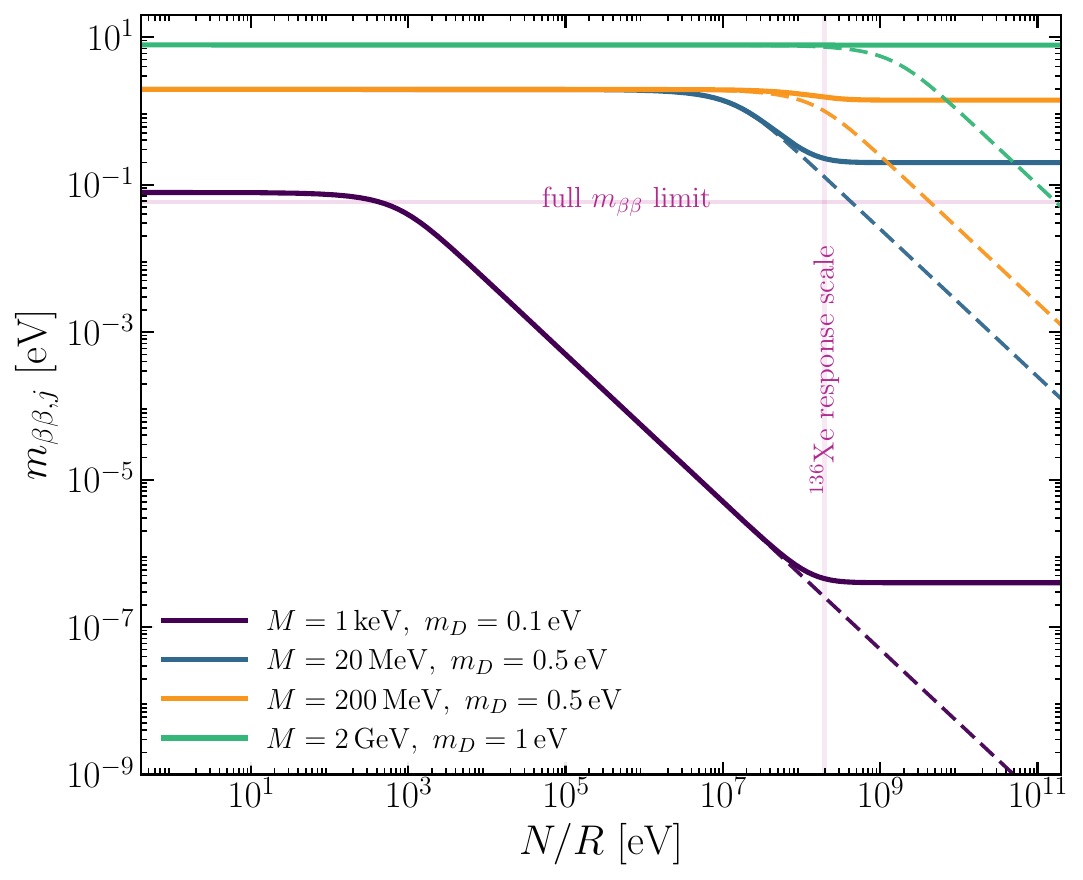}
    \caption{KK screening of the effective Majorana mass as a function of the highest included KK mode $N/R$, for $\mu_1=0.4~\mathrm{eV}$. Solid lines show $m_{\beta\beta,j}(N,p)$ as defined in Eq.~\eqref{eq:mbb_general}. The dashed lines show the unweighted partial sum $S(N)$ of Eq.~\eqref{eq:SN-def}. The purple solid lines show, for reference only, the matched KamLAND--Zen limit (horizontal) and the Argonne and CD--Bonn response scales (vertical).
    }
    \label{fig:screening-benchmarks}
\end{figure}
\bigskip
\paragraph{Kaluza-Klein Screening.}
In the region of interest for the DD scenario, the parameters follow the hierarchy $p\gg\mu_1,M_j\gg m_{D,j}$. In this limit Eq.~\eqref{eq:mbb_general} becomes
\begin{equation}
m_{\beta\beta}\approx\left|\sum_{j=1}^3U_{ej}^2\,\frac{\pi m_{D,j}^2}{\mu_1}\,\frac{M_j}{p}\right|\,,
\label{eq:0nubb-hierarchy}
\end{equation}
which is suppressed by an extra factor $M_j/p$ compared to the typical seesaw expectation.
A similar Kaluza-Klein suppression in $m_{\beta\beta}$ was pointed out in Ref.~\cite{Bhattacharyya:2002vf}.
The suppression of Eq.~\eqref{eq:0nubb-hierarchy} has its origin in gauge protection.
Gauge invariance forces the sum rule
\begin{equation}
    \label{eq:gauge-protection}\sum_m (V^j_{0m}\big)^2\,\lambda_{j,m}=0\,.
\end{equation}
for each generation $j$.
In the regime $p\gg\mu_1,M_j$, the interpolation factor of Eq.~\eqref{eq:mbb_spectral} approaches unity, and all the light states with mass $\lambda_{j,m}\ll p$ efficiently participate via Eq.~\eqref{eq:gauge-protection} in suppressing $m_{\beta\beta}$.
The suppression is somewhat analogous to the GIM mechanism in flavour physics and was already noted in the literature in models with light Majorana neutrinos~\cite{Blennow:2010th}.
In the case of KK Majorana neutrinos, such an effect is brought to its extreme.

In order to better understand this mechanism and understand the scaling of the screening, let us focus on a single generation and parametrise the truncated version of Eq.~\eqref{eq:gauge-protection} via
\begin{equation}
\label{eq:SN-def}
S(N)\equiv \sum_{n=0}^{N}\Big[\mathcal N_{+n}^2\lambda_{+n}+\mathcal N_{-n}^2\lambda_{-n}\Big]\,,
\end{equation}
so that $m_{\beta\beta}\simeq S(N)$, where with the $\pm 0$ subscript we indicate the lightest mode and its partner, and $N$ is an arbitrary KK-number cut-off. The sum at leading order in the expansion in $m_D$ and $1/N$ can be computed exactly and gives (cf.~App.~\ref{app:KKM-screening})
\begin{align}
&S(N)\approx \frac{2m_D^2 M}{\mu_1^2 N}\,.
\label{eq:screening-large-N}
\end{align}
As the number of states $N$ increases, the partial sum becomes more suppressed by $1/N$, and the KK tower screens the effective Majorana mass. The partial sum can be reconnected to $0\nu\beta\beta$:
$N$ is fixed by the interpolation coefficient to be $\lambda_N\simeq \mu_N\simeq p$, giving $S(p)=2m_D^2M/(\mu_1 p)$. This is up to $\mathcal{O}(1)$ corrections exactly  the expression of $m_{\beta\beta}$ derived in Eq.~\eqref{eq:0nubb-hierarchy} in the $p\gg M,\mu_1$ regime.

The screening mechanism is illustrated in Fig.~\ref{fig:screening-benchmarks}. We fix $\mu_1=0.4~\mathrm{eV}$ and show four benchmarks: $M=1$~keV and $20$~MeV in the screened regime, $M=200$~MeV across the nuclear-response transition, and $M=2$~GeV in the effectively unscreened regime. The values of $m_D$ are chosen among the larger values considered in our analysis, for which $0\nu\beta\beta$ can become relevant. The solid curves show the corresponding single-tower effective mass $m_{\beta\beta,j}(N,p)$, including the nuclear response of Eq.~\eqref{eq:mbb-interpolation}. The dashed curves show the gauge partial sum $S(N)$ defined in Eq.~\eqref{eq:SN-def}, which decreases as $1/N$ as additional KK pairs participate in the cancellation.
Once the KK mass exceeds the nuclear scale dictated by $p$, heavier modes become suppressed, and the result freezes, giving rise to the horizontal plateau in the figure.
As can be seen, for $M\ll p$, the light-mode contribution can lie close to the present experimental sensitivity, but the large number of KK modes below the nuclear scale screens the physical amplitude by several orders of magnitude. For $M\sim p$, only a partial cancellation takes place. Finally, for $M\gg p$, the nuclear response strongly suppresses the heavy KK states. They therefore cannot cancel the light-mode contribution, and the amplitude approaches the light-mode result.

\bigskip
\paragraph{Qualitative Constraint.}
\label{sec:mbb-phase-envelope}
To set a qualitative constraint, we take the expression of Eq.~\eqref{eq:0nubb-hierarchy} and focus on a single term of the sum, ignoring the PMNS factors and possible Majorana phases.
In the screened regime relevant for the DD, $p\gg M,\mu_1,m_D$, we find
\begin{equation}
  \mu_1\gtrsim  \frac{\pi m_D^2 M}{m_{\beta\beta}^{\rm lim}p}\propto M\,.
\label{eq:scaling-0nubb}
\end{equation}
Therefore, one expects the constraint on $R$ obtained from the upper envelope to become stronger as the Majorana mass increases.
Notice that the presence of Majorana phases in Eq.~\eqref{eq:mbb_general} can introduce cancellations among the terms of the sum: the true, phase-minimised constraint can be significantly weaker.
\subsection{Tritium Beta Decay}
\label{sec:tritium}
\paragraph{Observable.}
The last laboratory probe we include in the analysis is tritium beta decay: 
\begin{equation}
    ^3\text{H}\to\,^3\text{He} + e^- + \bar\nu_e\,.
\end{equation}
The non-vanishing neutrino masses lead to a distortion of the electron energy spectrum in beta decay, which can in turn be measured to set an upper bound on the neutrino masses.
Consequently, the observable is insensitive to the Dirac or Majorana nature of the neutrinos, as it can only probe the kinematics of the process.

Since the electron neutrino is not a mass eigenstate but a superposition of the physical
states, the decay kinematics is sensitive to the masses $m_j$ weighted by the mixing $|\mathbf{U}_{ej}|^2$. For light neutrinos, the electron effective mass is defined by
\begin{equation}
    m_\beta^2 \equiv \sum\limits_{j=1}^3 m_j^2 |\mathbf{U}_{ej}|^2 \,.
\end{equation}
The current strongest constraint comes from the KATRIN experiment~\cite{KATRIN:2024cdt}
\begin{align}
    \label{eq:mbeta-bound}&m_\beta \leq 0.45~\text{eV} \,, &90\%~\text{C.L.}
\end{align}
Compared to $0\nu\beta\beta$, the study of this observable is far more convoluted.
If additional mass eigenstates lie within the measured energy window, they can generate additional kinematic thresholds or kink-like spectral distortions. Depending on their mass spacing relative to the experimental resolution, these features may be individually resolved or may combine into an approximately continuous deformation of the spectrum, requiring a dedicated multi-state spectral analysis~\cite{ McLaughlin:2000iq,Basto-Gonzalez:2012nel,KATRIN:2025lph,Antoniadis:2025rck}.
However, for part of the parameter space of interest we can apply a useful simplification. The bound of Eq.~\eqref{eq:mbeta-bound} is derived from an endpoint-spectrum fit in which the signal normalisation, endpoint, and background are free parameters~\cite{Kleesiek:2018mel,KATRIN:2024cdt}. If the non-light threshold of every tower lies above the fitted window (approximately $M_j\gtrsim40$~eV in the weak-mixing regime), and the retained light roots lie inside it, the appropriate effective mass is~\cite{Farzan:2002zq,Rodejohann:2014eka,Kleesiek:2018mel,Antoniadis:2025rck}
\begin{align}
 m_{\beta,\mathrm{eff}}^2&\simeq \frac{\sum_{j=1}^3|U_{ej}|^2|V^j_{00}|^2\lambda_{j,0}^2}{\sum_{j=1}^3|U_{ej}|^2|V^j_{00}|^2}\,.
\label{eq:mb_spectral}
\end{align}
This approximation fails when a KK mode enters the fitted window, or if the lightest mode itself is displaced beyond it.  Those cases require the unexpanded multi-threshold beta spectrum (and, for a large common endpoint displacement, the independent tritium $Q$ value), rather than a direct application of Eq.~\eqref{eq:mbeta-bound}.

\bigskip
\paragraph{Qualitative Constraint.}
Finally, we turn to the qualitative scaling of the beta-decay constraint.
In the area of interest, $m_{D,j}$ is the smallest scale in the problem; then the lightest eigenstate satisfies $|V_{00}^j|^2\approx1$, and Eq.~\eqref{eq:mb_spectral} simplifies to
\begin{align}
|U_{ej}V_{00}^j|^2\,\lambda_{j,0}^2 &\;\approx\; |U_{ej}|^2\,\frac{\pi^2 m_{D,j}^4}{\mu_1^2}\,\coth^2\!\left(\frac{\pi M_j}{\mu_1}\right)\,.
\label{eq:lightest-mb}
\end{align}
Taking into account the asymptotic $\coth(x)\to1/x$ as $x\to0$ and $\coth(x)\to1$ as $x\to\infty$, the factor is proportional to $m_{D,j}^4/M_j^2$ for $\mu_1\gg M_j$ and to $m_{D,j}^4/\mu_1^2$ if vice versa. All in all, Eq.~\eqref{eq:lightest-mb} scales inversely with the square of $\min\{M,\mu_1\}$.
In the weak-mixing regime, the
KATRIN limit gives the useful estimate
\begin{equation}
 m_D\lesssim
 \left[\frac{(0.45~\mathrm{eV})\mu_1}{\pi}
 \tanh\!\left(\frac{\pi M}{\mu_1}\right)\right]^{1/2}.
\label{eq:mD-guidance-beta}
\end{equation}
In the limit relevant for the DD where $M\gg \mu_1$, the expression simplifies and the dependence on $M$ disappears. Comparing with the expression of the lightest mode of Eq.~\eqref{eq:lightest}, the KATRIN bound then translates to the usual bound on the absolute neutrino-mass scale, blind to any information about the Majorana mass or the extra-dimensional nature of the problem. The extra dimension enters only when translating this constraint into the model parameters, $m_D$ and the brane Yukawa coupling of Eq.~\eqref{eq:mD-yukawa-DD}, whose relation to the light mass depends on $\mu_1$ (cf.~Sec.~\ref{sec:results}).

All in all, one expects a flat constraint on $\mu_1$ as a function of the Majorana mass.

\bigskip
\paragraph{Future scope}
KK states with $M\sim1$--$10$~keV lie below the tritium endpoint $Q\simeq18.6$~keV and can in principle be produced directly. The upcoming TRISTAN upgrade of KATRIN~\cite{KATRIN:2018oow} targets exactly this regime, measuring the full tritium spectrum differentially in search of keV-scale kinks, with a projected sensitivity $|U_{ej}|^2(\mathcal N_{j,+n}^2+\mathcal N_{j,-n}^2)\sim10^{-6}$ for $D_{j,n}$ in the keV range~\cite{KATRIN:2026ywk}. How this translates into a bound on $m_{D,j}$ depends on whether individual KK levels are resolved, i.e.\ on $\mu_1$ relative to TRISTAN's energy resolution $\Delta E_{\rm TRISTAN}\sim300$~eV. If $\mu_1\gtrsim\Delta E_{\rm TRISTAN}$, each level appears as a separate kink, and using the deep-tower limit $\mathcal N_{j,+n}^2+\mathcal N_{j,-n}^2\approx2m_{D,j}^2/D_n^2$ gives
\begin{equation}
m_{D,j}\;\lesssim\;D_n\sqrt{\frac{10^{-6}}{2|U_{ej}|^2}}\,.
\label{eq:tristan-resolved}
\end{equation}
In the DD window, however, $\mu_1\lesssim2$~eV, far below $\Delta E_{\rm TRISTAN}$: neighbouring levels are not individually resolved, and the tower instead appears as a quasi-continuous distortion of the spectrum setting in at a neutrino mass $\simeq M$, rather than as discrete kinks. Coherently summing the $N_{\rm bin}\sim\Delta E_{\rm TRISTAN}/\mu_1$ nearly-degenerate levels within one resolution bin tightens the reach to
\begin{equation}
m_{D,j}\;\lesssim\;D_{j,n}\sqrt{\frac{10^{-6}\,\mu_1}{2|U_{ej}|^2\,\Delta E_{\rm TRISTAN}}}\,.
\label{eq:tristan-unresolved}
\end{equation}
Compared to the light-mode bound in Eq.~\eqref{eq:mD-guidance-beta}, both Eq.~\eqref{eq:tristan-resolved} and Eq.~\eqref{eq:tristan-unresolved} are stronger throughout most of $M\sim1$--$10$~keV, with an enhancement up to more than an order of magnitude at the low-$M$, large-$\mu_1$ end. Note that these considerations are purely qualitative, and a dedicated analysis of the continuum-like distortion of the spectrum is needed to draw constraints.

\section{Analysis and Benchmarks}
\label{sec:analysis}
We scan the compactification scale $\mu_1=1/R$ in the region allowed from fifth-forces searches of Eq.~\eqref{eq:fifth-forces} $[10^{-2},4\times10^2]$~eV  and the reference Majorana mass within $[10^{-3},10^{10}]$~eV. For NO we define the free Dirac mass parameter as $m_D=m_{D,1}$; for IO we use $m_D=m_{D,3}$. At every scan point, the other two Dirac masses are determined from the exact light roots so as to reproduce the measured solar and atmospheric mass splittings~\cite{Esteban:2024eli}. 
We consider three benchmarks for the bulk Majorana masses:
\begin{itemize}
    \item A \emph{universal} benchmark, $M_1=M_2=M_3$. This scenario is perhaps too strict, but it is pedagogical and can capture the extreme case of naturalness.
    \item A \emph{profiled} benchmark, where one Majorana mass is kept as a free parameter, and the other two are varied independently over
    \begin{equation}
     M_j^{\rm profiled}\in [0.1,10]\times M^{\rm reference}\,.
    \end{equation}
    The $\pm1$-decade interval is a naturalness assumption to remain consistent with the DD scenario: widening it can only weaken the resulting constraints. The free Majorana mass belongs to the reference generation ($M_1$ for NO and $M_3$ for IO). 
    \item Finally, the \emph{hierarchical} benchmark, which is the limiting case of Hierarchical Majorana masses. In this case, only one $M_j$ is appreciable, and the other two are fixed to $10^{-9}$~eV, a numerically stable representation of the lepton-number-conserving limit. This limit \textit{does not} sit within the DD framework. Nevertheless, it is a useful study to quantify possible deviations when some of the previous working assumptions are relaxed.
\end{itemize}
For our oscillation analysis, we adopt a spectral recast of the 3158-day Daya Bay data release~\cite{DayaBay:2022orm}. We combine the six-, eight-, and seven-detector data-taking periods using the 26 published far-hall prompt-energy bins over the range $0.7$--$12$~MeV. The predicted spectrum is averaged over the reactor--detector baselines, weighted by the corresponding target masses and detection efficiencies consistently with the released spectra. We use a multivariate Gaussian likelihood using the published bin-to-bin correlation matrix and relative systematic uncertainties. The experimental data have also been compiled in~Ref.~\cite{Eller:2026icp}

For each ordering, we define
\begin{equation}
 \Delta\chi^2_{\rm DB}=-2\ln\mathcal L_{\rm KK}
 +2\ln\mathcal L_{3\nu}
\end{equation}
relative to the three-neutrino prediction of the same ordering, drawing the contour at 90\% C.L.  The standard oscillation parameters are fixed to the NuFit 6.0 global-fit inputs~\cite{Esteban:2024eli}.
It must also be noted that Daya Bay operates on anti-electron modes and for $\nu_e\to\nu_e$ disappearance the weights $|U_{e i}|^2 (V_{0i}^n)^2$ are dominated by the $i=1,2$ components because $|U_{e3}|\ll |U_{e1}|,|U_{e2}|$. This in turn results in a stronger bound on the Inverted Ordering spectrum.
For the computation of the oscillation probability, we always include enough modes to saturate the unitarity condition $\sum_\lambda \mathcal{N}_\lambda^2=1$ up to $99\%$. 

\begin{figure*}[t]
  \begin{subfigure}[t]{0.495\textwidth}
    \centering
    \centering
    \includegraphics[width=\textwidth]{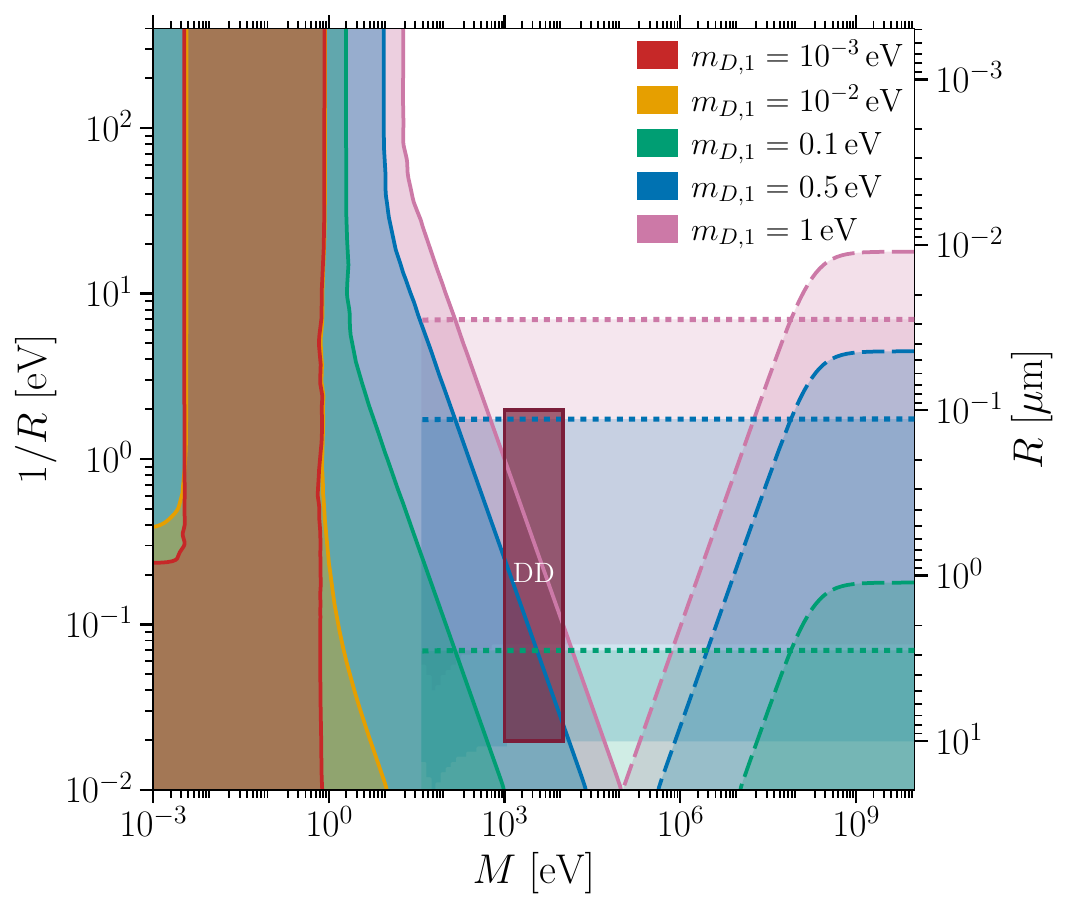}
    \caption{NO: Universal.}
    \label{fig:main-NO-universal}
  \end{subfigure}
  \begin{subfigure}[t]{0.495\textwidth}
    \centering
    \centering
    \includegraphics[width=\textwidth]{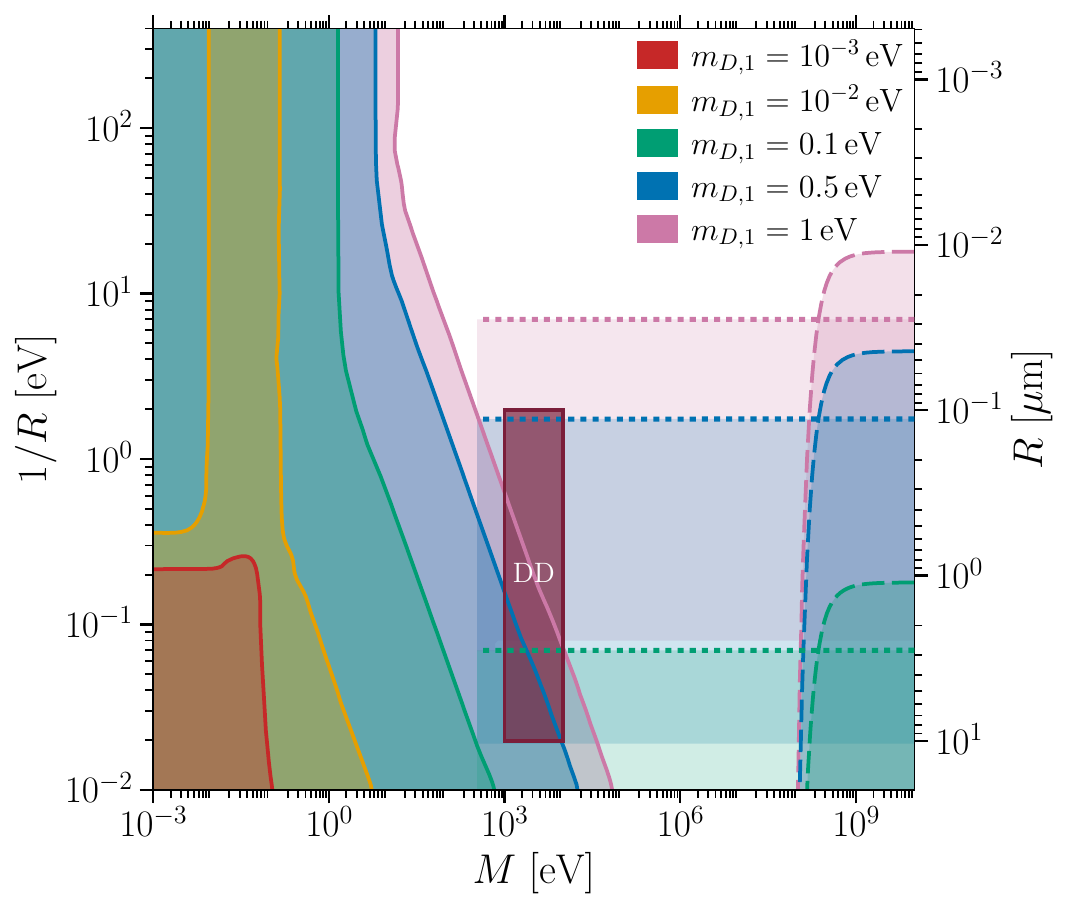}
    \caption{NO: Profiled.}
    \label{fig:main-NO-profiled}
  \end{subfigure}
    \caption{Exclusions at 90\% C.L. in the $(M,1/R)$ plane for NO with universal Majorana masses (left) and with the other two masses profiled over $M_{2,3}/M_1\in[0.1,10]$ (right). Colour denotes $m_{D,1}=10^{-3},10^{-2},0.1,0.5,$ and $1$~eV. Solid contours show the Daya Bay recast, dashed contours the conservative KamLAND-Zen $^{136}$Xe response envelope, and dotted contours the approximate KATRIN reinterpretation. The latter is drawn only for $M>40$~eV in the universal case and $M_1>400$~eV in the profiled case, ensuring that every non-light threshold lies above the analysis window in which we trust the approximation. The maroon rectangle marks the Dark-Dimension window, which should only be understood as an indicative guide. Notice that the profiled contours of panel (b) are not converged: the minimum over the nuisance masses frequently lies at the boundary of the sampled window. They should not be read as fully marginalised likelihood bounds (cf.~Sec.~\ref{sec:results-profiled}).}
    \label{fig:main-NO}
\end{figure*}

\bigskip
For the $0\nu\beta\beta$ constraint, at each point we evaluate the exactly resummed KK amplitude of Eq.~\eqref{eq:mbb_general} within each of the two matched $^{136}$Xe response interpolations described in Sec.~\ref{sec:0nubb}.
Because $m_{\beta\beta}$ is a sum of three terms with independent Majorana phases, its value ranges between a phase-aligned maximum and a phase-cancelled minimum. The aligned-phase (maximum) configuration $a_j\equiv |U_{ej}|^2m_{\beta\beta,j}$  gives the upper value
\begin{equation}
 m_{\beta\beta}^{\max}=\sum_j a_j=\sum_j|U_{ej}|^2\left|\mathrm{Re}\!\left[-\frac{p^2}{f^{(j)}(ip)}\right]\right|\,.
\label{eq:mbb-upper-envelope}
\end{equation}
On the contrary, in the worst scenario, free relative phases can be made to cancel down to
\begin{equation}
 m_{\beta\beta}^{\min}=\max\Big(0,\ 2\max_j a_j-\textstyle\sum_j a_j\Big)\,.
 \label{eq:mbb-lower-envelope}
\end{equation}
We present the \textit{most conservative} bound: we adopt the Argonne $m_{\beta\beta}^{\rm lim}$ limit from Tab.~\ref{tab:mbb-values}, and we consider a point to be excluded only if the most favourable (phase-cancelled) configuration is already ruled out, i.e.\ if $m_{\beta\beta}^{\min}>m_{\beta\beta}^{\rm lim}$.
Notice that we do not include other theory uncertainties such as quenched-$g_A$ variants or the  $\sim 20-25\%$ interpolation error near $m_j\sim p$~\cite{Faessler:2014kka}. The contours use the published KamLAND--Zen half-life and are not a reanalysis of its event likelihood.

\bigskip
For KATRIN, we use Eq.~\eqref{eq:mb_spectral} and the published $m_\beta<0.45$~eV limit only when every non-light threshold lies above the approximately $40$~eV fit window, and every retained light root lies inside it. We take a conservative approach and include the constraint only when such a condition is met. This requires $M>40$~eV for universal masses and $M^{\rm ref}>400$~eV in the profiled benchmark. The resulting dotted contours are limited to such a range of validity as they do not stem from a KATRIN spectral refit, which would arguably strengthen the constraint.

\section{Results and Discussion}
\label{sec:results}
We present and discuss the results separately for the three benchmarks: universal, profiled, and hierarchical Majorana masses. The first two are chosen consistently with the naturalness assumptions of the DD model. The hierarchical benchmark is studied only for completeness; we discuss it separately in the main text, and present the exclusion plots in App.~\ref{app:hierarchical}.

Before proceeding, it is instructive to anticipate some of the results in the regime $M\gg\mu_1$, favoured by the DD scenario. As we shall see, the dominant constraints arise from the combination of neutrino oscillation data and KATRIN, while in this region neutrinoless double beta constraints are suppressed by the KK-screening of Eq.~\eqref{eq:0nubb-hierarchy}. Interestingly, using the expression for the lightest mode eigenvalue of Eq.~\eqref{eq:lightest}, one finds that these constraints are effectively insensitive to the extra-dimensional nature of the model. Indeed, the KATRIN constraint of Eq.~\eqref{eq:mD-guidance-beta} reduces to the standard upper bound on the absolute neutrino-mass scale, $\lambda_{j,0}\lesssim0.45$~eV, which consequently constrains the largest Dirac Yukawa coupling.
Similarly, oscillation data are sensitive to the lightest modes of the KK towers, but impose a lower bound on the same mass scale, since the heaviest light neutrino must reproduce the largest measured mass splitting, $\lambda_{j,0}\gtrsim(\Delta m^2_{\max})^{1/2}\simeq0.05$~eV~\cite{Esteban:2024eli}. For the largest Dirac mass, the resulting window, $0.05\lesssim\lambda_{j,0}\lesssim0.45$~eV, translates into
\begin{equation}
0.13~\mathrm{eV}\lesssim m_{D,\max}\left(\frac{\mu_1}{\mathrm{eV}}\right)^{-1/2}\lesssim0.38~\mathrm{eV}\,.
\label{eq:mD-window}
\end{equation}
Using the relation between $m_D$ and the Yukawa parameter $y_D$ given in Eq.~\eqref{eq:mD-yukawa-DD-natural}, we further obtain
\begin{equation}
2\times10^{-3}\lesssim y_{D,\max}\left(\frac{\mu_1}{\mathrm{eV}}\right)^{-1/6}\lesssim5\times10^{-3}\,,
\label{eq:y-estimation}
\end{equation}
with a weak dependence on the exact value of $\mu_1$.

In the qualitative treatment we are discussing here, these bounds apply to the largest Yukawa coupling, while their implications for the remaining couplings depend on the assumptions of the model. If the Yukawa couplings are allowed to be hierarchical, the lightest neutrino could be massless, and KATRIN provides the dominant constraint by setting an upper bound. Conversely, if the DD framework favours some degree of naturalness among the Yukawa couplings, avoiding large hierarchies, the lower bound from oscillation data also indirectly constrains the order of magnitude of the smallest coupling. All in all, the DD phenomenologically relevant window for $m_D$ corresponds to a small five-dimensional brane Yukawa coupling, $y_D\lesssim 5\times10^{-3}$.


\subsection{Universal Majorana Masses}
We begin by considering the most constrained case where all Majorana masses are equal.
The resulting constraints at $90\%$~C.L. are shown in Figs.~\ref{fig:main-NO-universal}-\ref{fig:main-IO-universal} for NO and IO, respectively.
The DD-favoured region is probed for $m_D\gtrsim0.1$~eV and is ruled out by the KATRIN reinterpretation for the $m_D=1$~eV slice. 

The solid Daya Bay contours provide complementary reach towards smaller $M$ and strengthen with $m_D$. Their asymptotic weakening with increasing $M$ follows the $1/M$ scaling of Eq.~\eqref{eq:scaling-oscillations}. Conversely, $0\nu\beta\beta$ does not set the leading constraint in the keV DD box as $M\ll p\simeq200$~MeV, so the gauge-protected KK sum screens the nuclear-scale amplitude approximately as $M/p$. The dashed contours become relevant only at substantially larger $M$ and use the phase-minimised criterion of Eq.~\eqref{eq:mbb-lower-envelope}.

NO and IO display the same qualitative division of sensitivity, although Daya Bay is somewhat stronger in IO because electron disappearance is dominated by the $i=1,2$ components while $|U_{e3}|$ is small.

\subsection{Profiled Majorana Masses}
\label{sec:results-profiled}

\begin{figure*}[t]
    \centering
    \begin{subfigure}[t]{0.495\textwidth}
    \centering
    \includegraphics[width=\columnwidth]{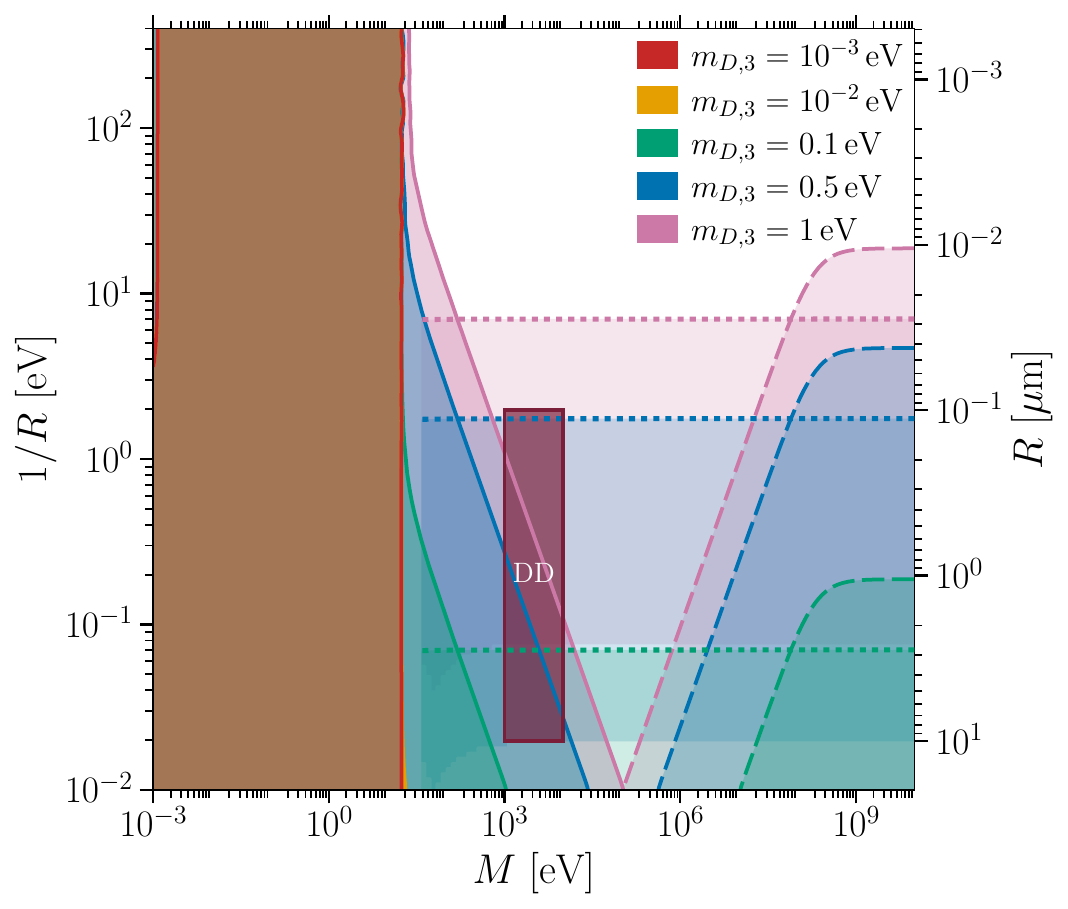}
    \caption{IO: Universal.}
    \label{fig:main-IO-universal}
    \end{subfigure}
    \centering
    \begin{subfigure}[t]{0.495\textwidth}
    \centering
    \includegraphics[width=\columnwidth]{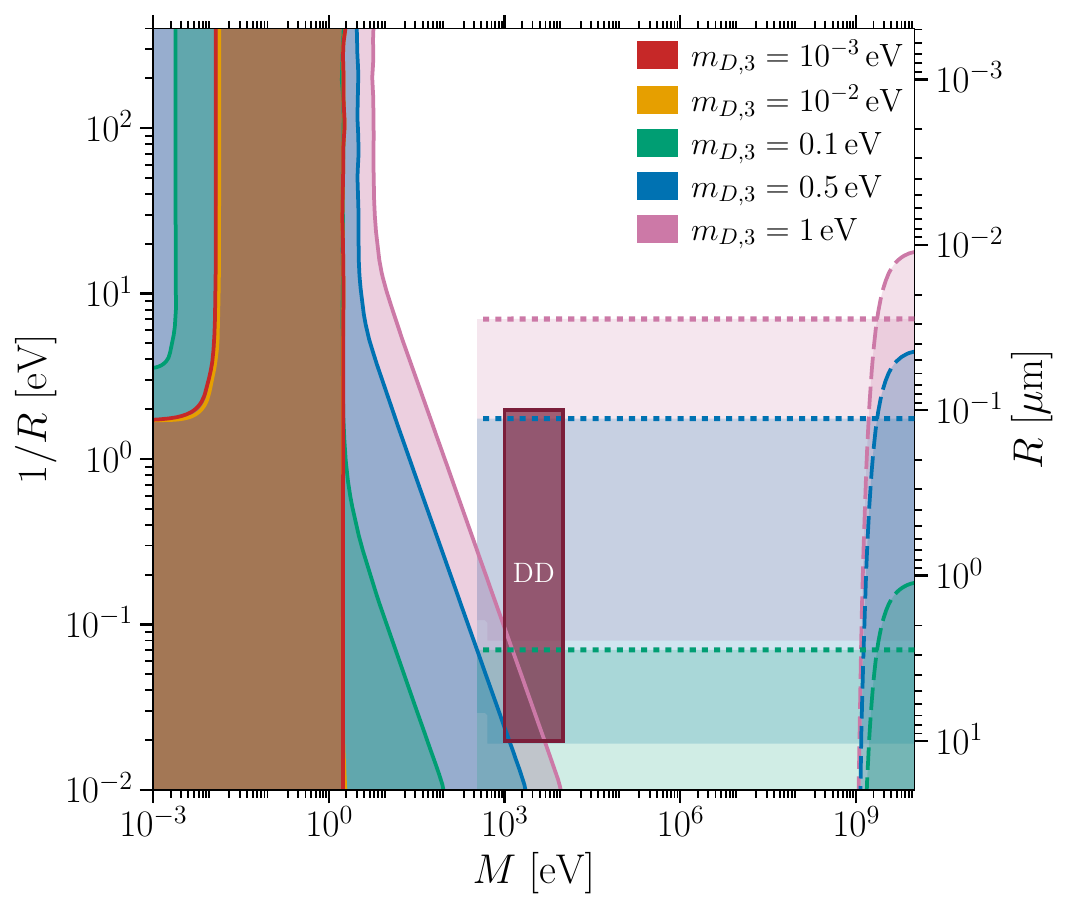}
    \caption{IO: Profiled.}
    \label{fig:main-IO-profiled}
    \end{subfigure}
    \caption{As in Fig.~\ref{fig:main-NO}, but for IO, with $m_D\equiv m_{D,3}$ and $M\equiv M_3$. The left panel assumes universal Majorana masses, and the right panel profiles the other two masses over the stated $\pm1$-decade window. As in Fig.~\ref{fig:main-NO}, the profiled contours of panel (b) are not converged, since the nuisance minimum frequently lies at the boundary of the sampled window.}
    \label{fig:main-IO}
\end{figure*}
The benchmark of the previous section is overly constraining as it is somewhat unlikely that Nature fixes three parameters to be identical.  To compensate, we repeat the analysis allowing for a small hierarchy among them, whose relative range is comparable to the allowed range for $R$ and $M$.  More specifically, we choose one of them as a free parameter, and allow the remaining two to vary within two orders of magnitude: one order of magnitude larger or smaller. However, we do not claim that this study constitutes a fully converged marginalised scan of the parameter space. In many cases, the minimum with respect to the nuisance parameters lies at the boundary of the sampled grid, preventing the resulting contours from being interpreted as robust profile-likelihood bounds. 

The results are shown in Figs.~\ref{fig:main-NO-profiled}-\ref{fig:main-IO-profiled} for NO and IO, respectively.
This freedom affects the probes differently. Across the displayed domain, the Daya Bay boundary changes comparatively little for the phenomenologically relevant larger-$m_D$ slices, whereas the robust $0\nu\beta\beta$ region can shrink substantially.
The contrast has a physical origin: the Daya Bay spectrum depends on disappearance probabilities and has no free Majorana phases, so changing the two nuisance masses mainly redistributes the KK frequencies and weights. The $0\nu\beta\beta$ amplitude is instead a coherent sum, and hence allowing independent $M_j$ values changes the three vector lengths $a_j$ and can make the triangle cancellation condition easier to satisfy. The discrete nuisance minimisation and the exact phase minimum are applied separately. At large $M$, this additional freedom can remove much of the remaining $0\nu\beta\beta$ exclusion, and the response also saturates once $M\gg p$ rather than continuing the small-$M$ linear scaling.

The profiled KATRIN contour is effectively insensitive to the profiling procedure. This is because we restrict the beta-decay analysis to the region of parameter space in which all heavier modes lie above $40$~eV, so they do not contribute within the relevant KATRIN energy window. Hence, profiling over these states has a negligible impact on the resulting contours.

\subsection{Hierarchical Majorana Masses}
The opposite limiting case has only one generation carrying an appreciable bulk Majorana mass. This is scenario does not fit within the Dark-Dimension proposal, but it isolates the flavour dependence and shows how strongly the conclusions rely on universality or naturalness assumptions. 

The six NO and IO panels are collected in~App.~\ref{app:hierarchical} in Figs~\ref{fig:onegen-NO}-\ref{fig:onegen-IO}.
For $0\nu\beta\beta$, the hierarchy is especially transparent. With only $M_j$ non-zero, the amplitude is proportional to the electron-flavour weight $|U_{ej}|^2$, and only a single $a_j$ is non-negligible, so $m_{\beta\beta}^{\min}=m_{\beta\beta}^{\max}=a_j$: the phase ambiguity of Eq.~\eqref{eq:mbb-lower-envelope} is irrelevant in this benchmark, and the upper-envelope estimate used previously is in fact exact here. 

The reach is therefore strongest for $j=1$, weaker for $j=2$, and strongly suppressed for $j=3$, reflecting the hierarchy of mixings $|U_{e1}|^2\simeq0.68$, $|U_{e2}|^2\simeq0.30$, and $|U_{e3}|^2\simeq0.02$. Inter-generation Majorana-phase cancellations are absent in this limit. Daya Bay remains sensitive because all three tower spectra enter electron-antineutrino disappearance, but the contour depends on which tower carries the Majorana splitting and on which reference Dirac mass is fixed by the ordering.

The one-generation panels also contain a sizeable no-solution region at low $\mu_1$. When two towers approach the Dirac limit, their light eigenvalues cannot exceed approximately $\mu_1/2$; below a generation-dependent threshold they cannot reproduce the measured mass splittings for any Dirac coupling, following directly from Eq.~\eqref{eq:spectrum-necessary}. We do not show a KATRIN contour in these panels because the quasi-Dirac spectator states lie inside its analysis window and require a dedicated multi-threshold spectrum fit.

\section{Conclusions}
\label{sec:conclusions}
Recent progress in the Swampland programme has singled out a ``Dark Dimension''~(DD) scenario. Such a framework proposes the existence of an extra mesoscopic spatial dimension within the $R\sim 0.1-10~\mu$m range equipped with three generations of bulk right-handed neutrinos with Majorana masses in the $M\sim 1-10$~keV range. Motivated by this, in this work, we derived general constraints on Majorana bulk neutrinos as a function of the Majorana masses and of the compactification radius. We then applied them to the DD-favoured parameter space to set constraints on the Yukawa couplings of the model, and assess the availability of such a scenario.

We analysed some of the most relevant neutrino laboratory observables, including flavour oscillation, neutrinoless double beta decay ($0\nu\beta\beta$) and beta decay. We analysed three benchmark scenarios: the case of universal Majorana masses, the case of profiled ones (i.e. where we let them vary within a fixed range) and a hierarchical case, where one of them is much larger than the others. The latter does not belong to the DD framework, but it is a useful benchmark to quantify the possible opening of the parameter space if some of the hypotheses on the DD are relaxed; we therefore report it for completeness in this work. 

The results are shown in Fig.~\ref{fig:main-NO} for Normal Ordering~(NO) and in Fig.~\ref{fig:main-IO} for Inverted Ordering~(IO), for the first two benchmarks. The plots for the hierarchical results can be found in App.~\ref{app:hierarchical}.
Our findings show that KATRIN provides the strongest constraint for the DD-favoured parameter space, while oscillations and $0\nu\beta\beta$ provide complementary constraints for $M\lesssim 1/R$ and $M\gg 1/R$ beyond the DD parameter space, respectively. In the DD-favoured region, the measured neutrino mass splittings and KATRIN impose lower and upper bounds, respectively, on the largest brane Yukawa coupling, placing it at $\mathcal{O}(10^{-3})$ (cf. Eq.~\eqref{eq:y-estimation}. This range depends only weakly on the compactification radius, scaling as $(1/R)^{1/6}$, and does not imply a comparable lower bound on each individual coupling.

It must be noted that the conjectures upon which this DD model is built cannot predict the exact range of parameters, neither for $R$ nor $M$. This leaves the possibility for the DD preferred region to fluctuate beyond the benchmark used in this work. Therefore, without exact coefficients, it is not possible to rule it out definitely.
Nevertheless, prospects may be able to increase the pressure on it. These include the TRISTAN update of KATRIN~\cite{KATRIN:2018oow}, which will be able to analyse the full beta spectrum beyond its endpoint, thus becoming sensitive to the Majorana mass dependence of the spectrum and strengthening the now leading constraints and extending the coverage over this scenario. Its ultimate reach will depend on how well the systematic uncertainties identified in Ref.~\cite{KATRIN:2026ywk} can be controlled.

In this work, we restricted ourselves to terrestrial experiments, thus neglecting cosmological observables. A thermally or gravitationally populated Kaluza--Klein tower can contribute to the effective number of relativistic species $N_\text{eff}$ and to the sum of neutrino masses. Indeed dedicated analyses suggest that these cosmological bounds can, in some regimes, be competitive with or stronger than the terrestrial ones discussed here~\cite{Vincent:2014rja,Anchordoqui:2024xvl, McKeen:2024fdl}. However, such bounds depend on the cosmological history of the extra dimension (reheating temperature, brane-bulk thermalisation, and the underlying cosmological model), which are largely independent assumptions from the ones entering the laboratory phenomenology. We therefore keep the two sets of constraints separate: the results presented here are, by construction, agnostic to the cosmological history of the Dark Dimension, while noting that the combination with cosmology is a natural and potentially very constraining direction for future work.

The future is bright for neutrino physics, and might be bright enough to shed light even on Nature's dark dimension.
\newpage
\subsection*{Acknowledgements}
We thank M.~Montero, C.~Vafa and I.~Valenzuela for engaging discussions which brought to the realisation of this paper.
We are also deeply grateful to M.~Bauer, E.~Fernandez-Martinez, I.~Martinez-Soler and J.~Turner for useful comments and suggestions which improved this work.
%
This article is based upon work from COST Action COSMIC WISPers CA21106,
supported by COST (European Cooperation in Science and Technology). 
Part of our numerical analyses were carried out with the help of Claude Code and Codex. All physical arguments, calculations, and conclusions were developed and verified by the authors, who take full responsibility for them.
\newpage
\appendix
\onecolumngrid
\begin{center}
  {\large \textbf{Appendix}
  }
\end{center}
\section{The Model and Spectrum}
\label{app:model-spectrum}
In this Appendix, we discuss in more detail the model setup and the derivation of the 4D EFT spectrum. We also discuss the structure of masses and mixings in interesting limits of the theory.
In this work, we work under the assumption that both the Dirac and the bulk Majorana mass matrices can be simultaneously diagonalised. This means that the problem of deriving the spectrum amounts to three copies of the same single-flavour one. For this reason, in the following, we omit flavour indices and always discuss a single flavour.
\subsection{5D Model}
\label{app:model}
The object of interest is the action of a bulk Majorana fermion $\Psi$ in flat spacetime,
\begin{equation}
S_{\Psi,\rm bulk} = \int d^4x\int_{-\pi R}^{\pi R} dy\,\left[
i\overline\Psi\Gamma^A\partial_A\Psi - \frac{M}{2}\overline\Psi\Psi^{5c}\hc\right]\,.
\label{eq-app:S5c}
\end{equation}
We adopt the following representation for the gamma matrices:
\begin{align}
    &\Gamma^\mu=\gamma^\mu\,, && \Gamma^5=i\gamma^5\,, &&\{\Gamma^A,\,\Gamma^B\}=2\eta^{AB}\,.
\end{align}
Throughout this work, we will employ the Weyl basis so that $\gamma^\mu$ can be written via $\sigma^\mu\equiv (1,\sigma^i)$ and $\bar{\sigma}^\mu\equiv (1,-\sigma^i)$, where $\sigma^i$ are the Pauli matrices. In particular, one has
\begin{align}
    & \gamma^\mu=\begin{pmatrix}
        0 & \sigma^\mu\\
        \bar{\sigma}^\mu & 0
    \end{pmatrix}\,, &&\gamma^5=\begin{pmatrix}
        -1 & 0\\
        0 & 1
    \end{pmatrix}\,.
\end{align}
In 5D it is not possible to construct an analogue of the $\gamma_5$, that is, an object that anticommutes with all five $\Gamma^A$. Therefore, no 5D chiral projectors can be consistently defined.
To go around this, chirality can be assigned by making use of the orbifold symmetry, which allows one to assign each field an orbifold parity
\begin{equation}
    \label{eq-app:parity5}\gamma^5 \Psi(x,-y)=\pm\Psi(x,y)\,.
\end{equation}
Without loss of generality, we choose to assign even orbifold parity ($+$) to $\Psi$. The field can then be decomposed in terms of Weyl spinors
such that $\Psi=\Psi_L+\Psi_R$
\begin{align}
    &\Psi=\begin{pmatrix}
        \psi_L\\
        \psi_R
    \end{pmatrix}\,, &&\Psi_L=\begin{pmatrix}
        \psi_L\\
        0
    \end{pmatrix}\,,
    &&\Psi_R=\begin{pmatrix}
        0\\
        \psi_R
    \end{pmatrix}\,.
\end{align}
The parity assignment of Eq.~\eqref{eq-app:parity5} translates into a parity for the chiral components
\begin{align}
   &\Psi_L(x,-y)=-\Psi_L(x,y)\,, &\Psi_R(x,-y)=+\Psi_R(x,y)\,.
\end{align}

The spinor $\Psi^{5c}$ appearing in Eq.~\eqref{eq-app:S5c} is the 5D Lorentz-preserving ``charge conjugation'' spinor.
The 5D charge conjugation matrix is defined by
\begin{equation}
C_5 \equiv \gamma^5 C\,,\qquad
\Psi^{5c} \equiv C_5\overline\Psi^T = \gamma^5\Psi^c\,,
\label{eq-app:C5def}
\end{equation}
where $C$ is the usual 4D charge conjugation operator
\begin{equation}
\Psi^c \equiv C\overline\Psi^T = C\gamma^0\Psi^\star\,,\qquad C=i\gamma^2\gamma^0\,,
\label{eq:C4d}
\end{equation}
The two charge conjugation matrices share some properties, but with an important difference. In this basis, the square of the operator gives $C_5^2=(\gamma^5C)^2=(\gamma^5)^2C^2=C^2=-\mathbb 1$, exactly like the ordinary 4D charge-conjugation matrix. What distinguishes them is instead the double conjugation of a spinor: using $\Psi^c=C\gamma^{0T}\Psi^\star$ and $\gamma^{0}C\gamma^0C=-\mathbb 1$ in the Weyl basis, one finds 
\begin{equation}
\big(\Psi^{5c}\big)^{5c}=C_5\gamma^{0T}\big(C_5\gamma^{0T}\Psi^\star\big)^\star = C_5\gamma^0C_5\gamma^0\,\Psi=-\Psi\,,
\label{eq-app:C5squared}
\end{equation}
using that $C_5$ is a real matrix in the Weyl basis and that $\gamma^0C_5\gamma^0=C_5$. This is to be contrasted with the ordinary 4D result $(\Psi^c)^c=+\Psi$. A single 5D bulk spinor therefore cannot satisfy the naive Majorana condition $\Psi^{5c}=\Psi$: only the combination entering the Lagrangian, $\overline\Psi\Psi^{5c}+\mathrm{h.c.}$, needs to be well defined, and it is this bilinear, rather than a pointwise reality condition on $\Psi$, that furnishes the genuinely 5D-Lorentz-invariant Majorana mass term used throughout this work.

\bigskip
Given the definitions above, the bulk action of Eq.~\eqref{eq-app:S5c} can be written in terms of its chiral components.
Since $\Psi^{5c}=\Psi_L^{5c}+\Psi_R^{5c}$, where, as established previously, $\Psi_L^{5c}$ transforms
as a 4D right-handed field and $\Psi_R^{5c}$ as a 4D left-handed field. The eigenvalues of $\gamma^5$ are $+1$ on right-handed and $-1$ on left-handed objects, so
\begin{equation}
\gamma^5\Psi_L^{5c} = +\Psi_L^{5c}\,,\qquad \gamma^5\Psi_R^{5c}=-\Psi_R^{5c}\,.
\end{equation}
Applying such a property and writing the spinor in terms of the 4D charge-conjugated spinors, we obtain\footnote{Notice the difference of a minus sign compared to what we would have obtained if we had defined the action of Eq.~\eqref{eq-app:S5c} with the 4D charge conjugation, which would have brought us to $\Psi^c=\Psi_L^c+\Psi_R^c$. }
\begin{equation}
\Psi^{5c} = \gamma^5\big(\Psi_L^c+\Psi_R^c\big) = \Psi_L^c - \Psi_R^c\,,
\label{eq:Psi5c}
\end{equation}
Writing $\overline\Psi=\overline\Psi_L+\overline\Psi_R$ and using Eq.~\eqref{eq:Psi5c}, we get
\begin{equation}
\overline\Psi\,\Psi^{5c} = \overline\Psi_L\Psi_L^c - \overline\Psi_R\Psi_R^c\,.
\label{eq:massbilinear5c}
\end{equation}
All in all, the bulk action can be written in terms of the chiral 4D charge-conjugated fields as
\begin{equation}
S_{\Psi,\rm bulk} = \int d^4x\int_{-\pi R}^{\pi R} dy\,\left[
i\overline\Psi\Gamma^A\partial_A\Psi
- \frac{M}{2}\Big(\overline\Psi_L\Psi_L^c-\overline\Psi_R\Psi_R^c\Big)+\mathrm{h.c.}\right]\,.
\label{eq-app:S5c_expanded}
\end{equation}
\subsection{Dimensional Reduction and 4D EFT}
Let us now derive the equation of motion.
Varying Eqs.~\eqref{eq-app:S5c}-\eqref{eq-app:S5c_expanded} gives
\begin{equation}
i\Gamma^A\partial_A\Psi = M\Psi^{5c} = M\big(\Psi_L^c-\Psi_R^c\big)\,.
\end{equation}
Projecting onto chiralities, the system of equations of motion becomes
\begin{align}
i\slashed{\partial}\Psi_R - \gamma^5\partial_5\Psi_L &= -M\Psi_R^c\,, \\
i\slashed{\partial}\Psi_L - \gamma^5\partial_5\Psi_R &= +M\Psi_L^c\,.
\label{eq-app:eom5c}
\end{align}
The 4D EFT can be obtained by performing the integral over the extra dimension. In order to derive $y$-profile of the bulk field, the $y$-dependence is separated by performing the KK expansion of the fields
\begin{align}
    \label{eq-app:KK-decomposition}&\Psi_L(x,y)=\frac{1}{\sqrt{V}}\sum\limits_{n=0}^\infty \Psi_{L,n}(x) \xi_n(y)\,, &\Psi_R(x,y)=\frac{1}{\sqrt{V}}\sum\limits_{n=0}^\infty \Psi_{R,n}(x) \chi_n(y)\,,
\end{align}
where $\{\xi_n(y)\}_n$ and $\{\chi_n(y)\}_n$ are sets of eigenfunctions (hereafter called ``wavefunctions"~(WFs)), \( n\in\mathbb{Z}_{\ge0} \) labels the KK excitation level, and $V=2\pi R$ is the volume of the extra dimension, whose inclusion is to ensure the correct normalisation of the WFs
\begin{equation}
\label{eq-app:normalisation}
    \frac{1}{V}\int\limits_{-\pi R}^{\pi R} dy\, \xi_{n}(y)\xi_{m}(y)=\frac{1}{V}\int\limits_{-\pi R}^{\pi R} dy\, \chi_{n}(y)\chi_{m}(y)=\delta_{nm}\,.
\end{equation}
Compared to the flat case without Majorana mass, $M$ is constant along $y$, therefore, the bulk WFs are unaffected and have the same solutions~\cite{deGiorgi:2025xgp}:
\begin{align}
    &\xi_0(y)=0\,, &&\chi_0(y)=1\,, &&\xi_{n>0}(y)=\sqrt2\sin(\mu_n y)\,,
    &&\chi_{n>0}(y)=\sqrt2\cos(\mu_n y)\,, && \mu_n=n/R\,.
\end{align}
The WFs so obtained clearly satisfy the orbifold parity described above.
Substituting into
Eq.~\eqref{eq-app:S5c_expanded}, the kinetic term produces the KK Dirac mass tower as
for the flat case, so that the mass sector reads
\begin{equation}
S^\Psi_{\rm free} \supset -\int d^4x\left[
\sum_{n=1}^N \mu_n\,\overline\psi_{L,n}\psi_{R,n}
+ \frac{M}{2}\sum_{n=0}^N\Big(\overline\psi_{L,n}\psi_{L,n}^c
- \overline\psi_{R,n}\psi_{R,n}^c\Big) + \mathrm{h.c.}\right]\,.
\label{eq-app:Sfree5c}
\end{equation}
The above mass terms can be written in a more useful form. Starting from Eq.~\eqref{eq-app:Sfree5c} we perform a rotation writing
\begin{equation}
\psi_{R,n}=s_n\,\psi_{1,n}+c_n\,\psi_{2,n}\,,\qquad
\psi_{L,n}=c_n\,\psi_{1,n}^c-s_n\,\psi_{2,n}^c\,,\qquad n>0\,,
\label{eq-app:field_redef_5c}
\end{equation}
with the constant weights given by sines and cosines
\begin{equation}
s_n^2=\frac{D_n-M}{2D_n}\,,\qquad c_n^2=\frac{D_n+M}{2D_n}\,,\qquad
D_n\equiv\sqrt{M^2+\mu_n^2}\,.
\end{equation}
Substituting Eq.~\eqref{eq-app:field_redef_5c} into Eq.~\eqref{eq-app:Sfree5c},
the cross term $\overline{\psi_{1,n}^c}\psi_{2,n}$ cancels between the two rotated bilinears, leaving a purely diagonal result,
\begin{equation}
S \supset -\frac12\int d^4x\left\{
-M\,\overline\psi_{R,0}^c\psi_{R,0}
+ \sum_{n=1}^N\Big[D_n\,\overline\psi_{1,n}^c\psi_{1,n}
- D_n\,\overline\psi_{2,n}^c\psi_{2,n}\Big] + \mathrm{h.c.}\right\}\,,
\label{eq-app:Lfinal5c}
\end{equation}
The negative signs in front of the mass terms can be reabsorbed within the Majorana phases.

\subsection{Mixing with the SM}
\label{app:mixing-SM}
Coupling to the SM proceeds through the brane operator
\begin{equation}
    \mathcal{L}_\text{brane} \supset -\int d^4x\,m_D\,\overline\nu_L\Psi_R(x,0)+\mathrm{h.c.}\,.
\end{equation}
Expanding $\Psi_R(x,0)$ in KK modes with
$\chi_0(0)=1$, $\chi_{n>0}(0)=\sqrt2$, and using Eq.~\eqref{eq-app:field_redef_5c} yields
\begin{equation}
m_D\,\overline\nu_L\Psi_R(x,0) \;\longrightarrow\;
m_D\,\overline\nu_L\psi_{R,0}
+\sum_{n=1}^N\sqrt2\,m_D\Big(s_n\,\overline\nu_L\psi_{1,n}+c_n\,\overline\nu_L\psi_{2,n}\Big)\,,
\end{equation}
where we reabsorbed the $1/\sqrt{V}$ factor into the $m_D$ coupling.
In this basis, the brane coupling reaches $\psi_{1,n},\psi_{2,n}$ with level-dependent weights
$s_n,c_n$ rather than the uniform $m_D$ of the rotated construction. The $(\nu_L,\nu_L)$
entry is zero, protected by SM gauge symmetry. Collecting fields
into
\begin{equation}
X \equiv \big(\nu_L,\ \psi_{R,0}^c,\ \psi_{1,1}^c,\ \psi_{2,1}^c,\ \dots,\ \psi_{1,N}^c,\ \psi_{2,N}^c\big)^T\,,
\label{eq-app:Xvector5c}
\end{equation}
we can write the mass term in a more compact form $-\mathcal L\supset\tfrac12\overline X\,\mathbf M\,X^c+\mathrm{h.c.}$
with
\begin{equation}
\mathbf{M} =
\left(\begin{array}{ccccccc}
0 & m_D & \sqrt2m_Ds_1 & \sqrt2m_Dc_1 & \dots & \sqrt2m_Ds_N & \sqrt2m_Dc_N \\
m_D & -M & 0 & 0 & \dots & 0 & 0 \\
\sqrt2m_Ds_1 & 0 & D_1 & 0 & \dots & 0 & 0 \\
\sqrt2m_Dc_1 & 0 & 0 & -D_1 & \dots & 0 & 0 \\
\vdots & \vdots & \vdots & \vdots & \ddots & \vdots & \vdots \\
\sqrt2m_Ds_N & 0 & 0 & 0 & \dots & D_N & 0 \\
\sqrt2m_Dc_N & 0 & 0 & 0 & \dots & 0 & -D_N
\end{array}\right)\,.
\label{eq-app:massmatrix_final}
\end{equation}
The Majorana mass matrix $\mathbf M$ is real symmetric, hence automatically Hermitian: its eigenvalues and
eigenvectors are directly physical, up to the overall sign of each eigenvalue, which can be rotated away afterwards as a Majorana phase.
\subsection{Eigenvalues and Eigenvectors}
\label{app:eigensystem}
\begin{figure}
    \centering
    \includegraphics[width=\linewidth]{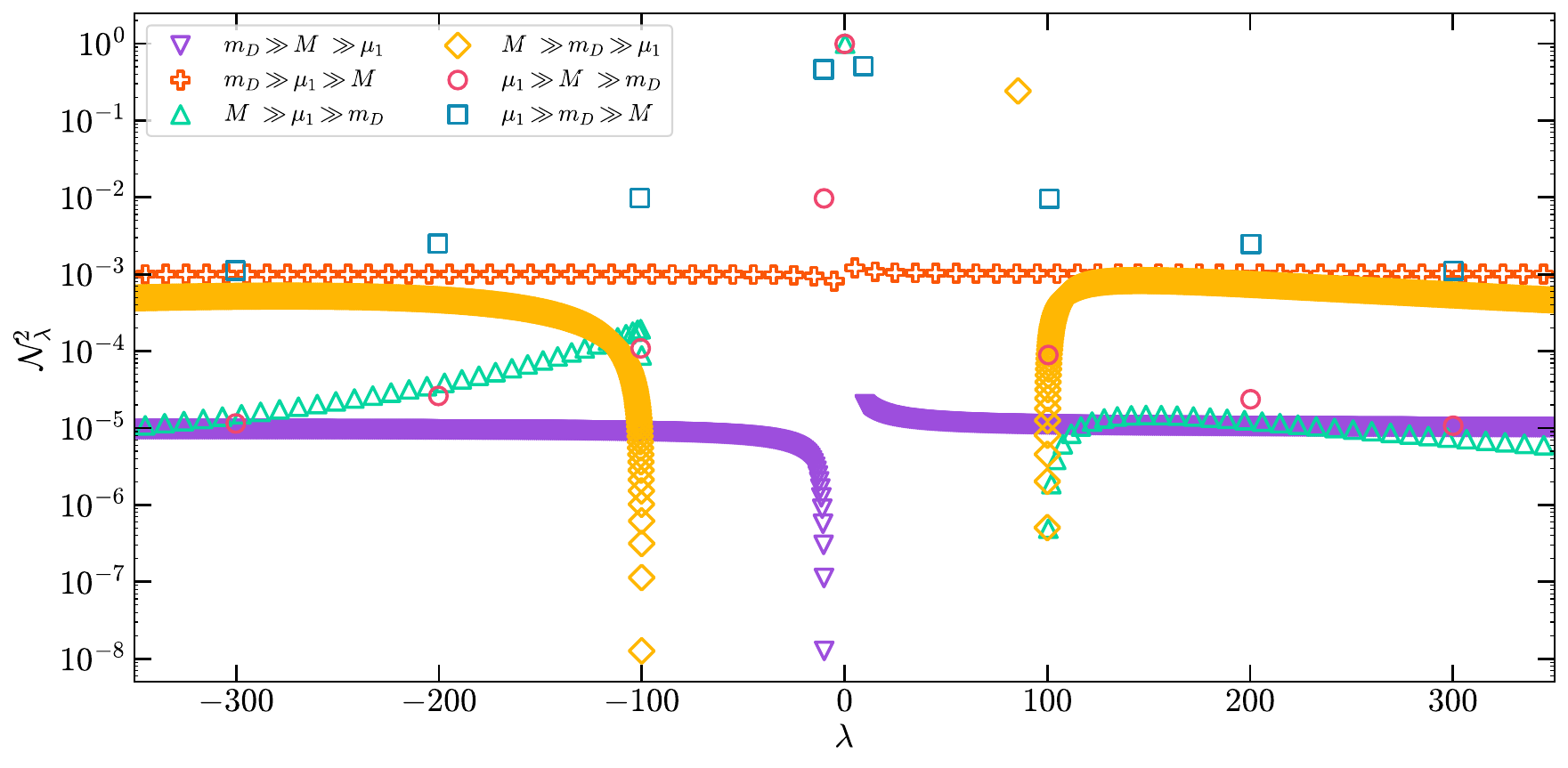}
    \caption{Masses and mixings of the 4D EFT for a representative set of hierarchical benchmark values in the format $X\gg Y \gg Z$, such that $X=100$, $Y=10$ and $Z=1$. All parameters are in arbitrary units.}
    \label{fig:spectra}
\end{figure}

We turn now to the derivation of the spectrum of the theory by computing the eigenvalues and eigenvectors of the Majorana mass matrix of Eq.~\eqref{eq-app:massmatrix_final}.
We can write $\mathbf M$ in block form as
\begin{equation}
\mathbf M = \begin{pmatrix} 0 & v^T \\ v & D \end{pmatrix}
\end{equation}
with
\begin{equation}
D = \mathrm{diag}\big(-M,\,D_1,\,-D_1,\,\dots,\,D_N,\,-D_N\big)\,,\quad
v=\big(m_D,\,\sqrt2m_Ds_1,\,\sqrt2m_Dc_1,\,\dots\big)\,.
\end{equation}
For $Mu=\lambda u$ with
$u=(u_0,u_1,\dots,u_{2N+1})^T$, the $i>0$ rows give $u_i=v_iu_0/(\lambda-D_{ii})$
(assuming $\lambda\neq D_{ii}$), and substituting into the $i=0$ row,
$\sum_iv_iu_i=\lambda u_0$, yields the secular function
\begin{equation}
f(\lambda)=\lambda -\sum\limits_{i>0}\frac{v_i^2}{\lambda-D_{ii}}\,,
\label{eq-app:secular_general}
\end{equation}
with corresponding eigenvector $u_\lambda\propto\big(1,\,v_1/(\lambda-D_{11}),\,v_2/(\lambda-D_{22}),\,\dots\big)^T$
for each root $\lambda$ of Eq.~\eqref{eq-app:secular_general}.
The eigenvalues are then determined as the roots of $f(\lambda)$.
We then get the equation and eigenvectors
\begin{equation}
\lambda = \frac{m_D^2}{\lambda+M}
+\sum_{n=1}^N\left[\frac{2m_D^2s_n^2}{\lambda-D_n}+\frac{2m_D^2c_n^2}{\lambda+D_n}\right]\,,
\qquad
u_\lambda = \mathcal N_\lambda\left(1,\ \frac{m_D}{\lambda+M},\
\frac{\sqrt2m_Ds_1}{\lambda-D_1},\ \frac{\sqrt2m_Dc_1}{\lambda+D_1},\ \dots\right)^T\,,
\label{eq-app:eigenvector_closed}
\end{equation}
where $N_\lambda$ is a normalisation coefficient.
Substituting $s_n^2,c_n^2$ from Eq.~\eqref{eq-app:field_redef_5c}, each bracket in the sum
collapses to a single term,
\begin{equation}
\frac{2m_D^2s_n^2}{\lambda-D_n}+\frac{2m_D^2c_n^2}{\lambda+D_n}
= \frac{2m_D^2(\lambda-M)}{w^2-\mu_n^2}\,,\qquad w\equiv\sqrt{\lambda^2-M^2}\,,
\end{equation}
so that
\begin{equation}
\lambda = \frac{m_D^2}{\lambda+M}+2m_D^2(\lambda-M)\sum_{n=1}^N\frac{1}{w^2-\mu_n^2}\,.
\end{equation}
As $N\to\infty$, the sum closes, leaving the closed-form transcendental equation
\begin{align}
&f(\lambda)=\lambda - \frac{\pi m_D^2(\lambda-M)}{\mu_1\,w}\,\cot\!\left(\frac{\pi w}{\mu_1}\right)=0\,,
& w=\sqrt{\lambda^2-M^2}\,.
\label{eq-app:eigenvalue_closed}
\end{align}
The normalization factor $\mathcal N_\lambda$ in Eq.~\eqref{eq-app:eigenvector_closed} can be computed similarly. We can simplify the problem by noticing that the norm of the eigenvector can be written in terms of the secular function $f(\lambda)$ of Eq.~\eqref{eq-app:secular_general}
\begin{equation}
\mathcal N_\lambda^{-2} = 1+\sum_i\frac{v_i^2}{(\lambda-D_{ii})^2} = f'(\lambda)\,,
\end{equation}
since $f'(\lambda)=1+\sum_i v_i^2/(\lambda-D_{ii})^2$ term by term, after some simplifications we obtain
\begin{equation}
\mathcal N_\lambda^{-2} = \frac{\lambda(\lambda-M)-M^2}{w^2}
+\frac{\pi^2m_D^2\,\lambda(\lambda-M)}{\mu_1^2\,w^2}\,
\csc^2\!\left(\frac{\pi w}{\mu_1}\right)\,,\qquad w=\sqrt{\lambda^2-M^2}\,,
\label{eq-app:normalization_closed}
\end{equation}
valid at each root $\lambda$ of Eq.~\eqref{eq-app:eigenvalue_closed}.
The masses and mixing with the SM neutrino for a representative set of benchmark values are shown in Fig.~\ref{fig:spectra}.
\section{Neutrinoless Double Beta Decay}
\label{app:0vbb}
Non-vanishing Majorana masses can in principle lead to neutrinoless double beta decay. Given the large number of states, the contribution could be relevant even for modest values of the Majorana mass. In this Appendix, we derive the relevant formulas for neutrinoless double beta decay, and show that the effect in the relevant parameter space is always suppressed compared to the standard Majorana sterile neutrino case.
We then specialize the exact three-flavour
in the regime in which the nuclear momentum scale $p\equiv\sqrt{\langle p^2\rangle}$
dominates over every scale in the KK construction, $p\gg \mu_1, M_{i}, m_{D,i}$
for $i=1,2,3$.
\subsection{Matched \texorpdfstring{$^{136}$Xe}{Xe-136} conversion}
\label{app:xe136-conversion}
For completeness, we give the numerical conversion used for the isotope-matched inputs in Sec.~\ref{sec:0nubb}. KamLAND--Zen uses $G^{0\nu}=14.54\times10^{-15}~\mathrm{yr}^{-1}$ and $g_A=1.269$~\cite{KamLAND-Zen:2024eml}, entering the dimensionally complete light-exchange relation
\begin{equation}
\left[T_{1/2}^{0\nu}\right]^{-1}=G^{0\nu}g_A^4|M_\nu|^2
\left|\frac{m_{\beta\beta}}{m_e}\right|^2\,,
\end{equation}
with $G^{0\nu}$ the phase space factor and $g_A$ the nucleon axial-vector coupling constant. With $T_{1/2}^{0\nu}=3.8\times10^{26}$~yr,
$\sqrt{G^{0\nu}T_{1/2}^{0\nu}}=2.3506\times10^6$. The Argonne value $M_\nu=2.177$ therefore gives
\begin{equation}
m_{\beta\beta}^{\rm lim}=
62.0~\mathrm{meV},
\end{equation}
while the CD--Bonn value $M_\nu=2.460$ gives $54.9$~meV. The corresponding heavy matrix elements, $M_N=152$ and $228$ ($M_N$ and $M_\nu$ being the nuclear matrix elements for the ``sterile'' and the light neutrinos, respectively), give the matched momenta $p=183.0$ and $210.8$~MeV through $p^2=m_em_p|M_N/M_\nu|$.

\subsection{Master Formula}
The quantity we are interested in is the electron-flavour Majorana mass. Using Eq.~\eqref{eq:basis-definition}, the full mass spectrum is labelled by the pair $(j,m)$, with eigenvalue $\lambda_{j,m}$ and electron-flavour overlap $U_{ej}V^j_{0m}$, so that (see also Refs.~\cite{Blennow:2010th,Abada:2018qok} for the analogous single-tower/light-sterile case)
\begin{equation}
m_{\beta\beta}=\left|\sum_{j}\sum_m \big(U_{e j}V^j_{0m}\big)^2\,\lambda_{j,m}\,\frac{\langle p^2\rangle}{\langle p^2\rangle+\lambda_{j,m}^2}\right|\,,
\label{eq-app:mbb_spectral}
\end{equation}
where $p\equiv\sqrt{\langle p^2\rangle}\sim \mathcal{O}(100)$~MeV is the associated nuclear momentum scale, and the double sum runs over every eigenstate of the full three-generation
KK system. Equation~\eqref{eq-app:mbb_spectral} extends the light/heavy
nuclear-matrix-element interpolation to the full tower
~\cite{Faessler:2014kka,Bolton:2022tds}.
The key point is noticing that $m_{\beta\beta}$ can be written in terms of the resolvent of $\mathbf M$,
\begin{equation}
G_{\alpha\beta}(z)\equiv\big[(\mathbf M-z\mathbb 1)^{-1}\big]_{\alpha\beta}=\sum_{j,m}\frac{\big(U_{\alpha j}V^j_{0m}\big)\big(U_{\beta j}V^j_{0m}\big)}{\lambda_{j,m}-z}\,,
\label{eq-app:Gdef}
\end{equation}
where the second equality follows from the spectral decomposition of $\mathbf M$ together with Eq.~\eqref{eq:basis-definition}. 
For convenience, we define $G_{00}^{(j)}(z)$ via
\begin{equation}
G_{\alpha\beta}(z)=\sum_j U_{\alpha j}U_{\beta j}\left(\sum_m\frac{\big(V^j_{0m}\big)^2}{\lambda_{j,m}-z}\right)\equiv\sum_{j=1}^3U_{\alpha j}U_{\beta j}\,G_{00}^{(j)}(z)\,.
\label{eq-app:Gee-split}
\end{equation}
The weight in Eq.~\eqref{eq-app:mbb_spectral} can be written as a sum of two simple poles
\begin{equation}
\frac{\lambda\langle p^2\rangle}{\langle p^2\rangle+\lambda^2}=\frac{\langle p^2\rangle}{2}\left[\frac1{\lambda-i p}+\frac1{\lambda+i p}\right]\,.
\end{equation}
Employing Eq.~\eqref{eq-app:Gee-split}, Eq.~\eqref{eq-app:mbb_spectral} collapses to
\begin{equation}
m_{\beta\beta}=p^2\left|\sum_{j=1}^3U_{ej}^2\,\mathrm{Re}\,G_{00}^{(j)}(ip)\right|\,,
\label{eq-app:mbb_p2ReG}
\end{equation}
where we used the reality of $\lambda_{j,m}$ and $V^j_{nm}$ to substitute $G_{00}^{(j)}(-ip)=G_{00}^{(j)}(ip)^\star$.

The whole problem is then reduced to computing each $G_{00}^{(j)}(ip)$, which contains the entire KK structure.
For each generation, $G_{00}^{(j)}(z)=\big[(\mathbf M^{(j)}-z\mathbb1)^{-1}\big]_{00}$, i.e.\ $(\nu_L,\nu_L)$, resolvent entry of the single decoupled tower $j$, computed directly in the interaction basis of $\mathbf M^{(j)}$.
Writing $\mathbf M^{(j)}$ in block form, $\mathbf M^{(j)}=\big(\begin{smallmatrix}0&v_j^T\\v_j&D_j\end{smallmatrix}\big)$, we extract $G_{00}^{(j)}(z)$ via the Schur complement of the lower-right block. Indeed for a generic block matrix $\mathbf{M}=\big(\begin{smallmatrix}A&B\\C&D\end{smallmatrix}\big)$ with $D$ invertible, it holds
\begin{equation}
[\mathbf{M}^{-1}]_{00}=\big(A-BD^{-1}C\big)^{-1}\,.
\end{equation}
Identifying $A=-z$, $B=v_j^T$, $C=v_j$, $D=D_j-z\mathbb1$ gives
\begin{equation}
G_{00}^{(j)}(z)=\Big[-z-v_j^T(D_j-z\mathbb1)^{-1}v_j\Big]^{-1}\,.
\end{equation}
Since $D_j$ is diagonal, $(D_j-z\mathbb1)^{-1}=1/((D_j)_{ii}-z)$, and thus
\begin{equation}
v_j^T(D_j-z\mathbb1)^{-1}v_j=\sum_i\frac{(v_j)_i^2}{(D_j)_{ii}-z}=-\sum_i\frac{(v_j)_i^2}{z-(D_j)_{ii}}\,,
\end{equation}
and therefore
\begin{equation}
-z-v_j^T(D_j-z\mathbb1)^{-1}v_j=-\left[z-\sum_i\frac{(v_j)_i^2}{z-(D_j)_{ii}}\right]=-f^{(j)}(z)\,,
\end{equation}
where we used the definition of $f^{(j)}$ from Eqs.~\eqref{eq-app:secular_general}-\eqref{eq-app:eigenvalue_closed}. This leads to the relation of interest
\begin{equation}
G_{00}^{(j)}(z)=\big[-f^{(j)}(z)\big]^{-1}=-\frac1{f^{(j)}(z)}\,.
\label{eq-app:G00_schur}
\end{equation}
Combining Eqs.~\eqref{eq-app:mbb_p2ReG} and \eqref{eq-app:G00_schur} gives the exact, closed-form, all-orders master formula
\begin{equation}
m_{\beta\beta}=p^2\left|\sum_{j=1}^3U_{ej}^2\,\mathrm{Re}\left[-\frac1{f^{(j)}(ip)}\right]\right|\,,
\label{eq-app:mbb_general}
\end{equation}
valid for any hierarchy among $p,\mu_1,M_j,m_{D,j}$ and for the full ($N\to\infty$) tower. 
As a consistency check, if we send $\mu_1 \to \infty$ we must recover the standard formula for three Majorana neutrinos. In such a limit, we find
\begin{equation}
    f^{(j)}(ip)=ip-\frac{m_{D,j}^2}{M_j+ip}\,,
\end{equation}
which leads to
\begin{equation}
    p^2\mathrm{Re}\left[-\frac1{f^{(j)}(ip)}\right]\approx \frac{M_j m_{D,j}^2 p^2}{M_j^2 p^2+\left(m_{D,j}^2+p^2\right)^2}\approx \frac{m_{D,j}^2}{M_j}\times \begin{cases}
           1\,, &M_j\gg p,m_{D,j}\,,\\
       \left(\frac{M_j}{p}\right)^2\,, &p\gg M_j,m_{D,j}\,.
    \end{cases}
\end{equation}
In the limit $M_j\gg p$, the quantity $m_{D,j}^2/M_j$ matches the neutrino mass, thus recovering the well-known Type-I seesaw result.
Similarly, in the opposite case where $p\gg M_j$, the result is further suppressed by additional powers of $M_j/p$, matching the literature once again~\cite{Abada:2018qok}.
\subsection{Kaluza-Klein Majorana Screening}
\label{app:KKM-screening}
We now specialize Eq.~\eqref{eq-app:mbb_general} to the regime of interest for the DD scenario: $p\gg\mu_1,M_j$. In this limit $w_j=\sqrt{-p^2-M_j^2}\to ip$ up to corrections of order $M_j^2/p^2$, and the argument of the cotangent in Eq.~\eqref{eq-app:eigenvalue_closed} becomes purely imaginary and parametrically large, $\pi w_j/\mu_1\to i\pi p/\mu_1\to i\infty$. Using $\cot(iy)=-i\coth(y)\to-i$ as $y\to\infty$, with corrections of order $e^{-2\pi p/\mu_1}$, Eq.~\eqref{eq-app:eigenvalue_closed} reduces to
\begin{equation}
f^{(j)}(ip)\simeq
i\left(p+\frac{\pi m_{D,j}^2}{\mu_1}\right)
-\frac{\pi m_{D,j}^2M_j}{\mu_1\,p}\,.
\end{equation}
Expanding $G_{00}^{(j)}(ip)=-1/f^{(j)}(ip)$ at leading order gives
\begin{equation}
\mathrm{Re}\,G_{00}^{(j)}(ip)\simeq\frac{\pi m_{D,j}^2M_j/(\mu_1p)}{\big(p+\pi m_{D,j}^2/\mu_1\big)^2}\simeq\frac{\pi m_{D,j}^2M_j}{\mu_1\,p^3}\,,
\end{equation}
which is manifestly positive.
Finally, substituting into Eq.~\eqref{eq-app:mbb_general} we find the result of interest
\begin{align}
&m_{\beta\beta}\;\simeq\;\frac{\pi}{\mu_1\,p}\left|\sum_{j=1}^3U_{ej}^2\,m_{D,j}^2\,M_j\right|\,, &p\gg\mu_1,\,M_j,m_{D,j}\,.
\label{eq-app:mbb_final}
\end{align}
For a single generation, this is simply $m_{\beta\beta}\simeq\pi m_D^2M/(\mu_1p)$: in the regime where the nuclear momentum scale dwarfs both new physics scales, the effective mass is set by the product of the Dirac-suppressed coupling and the bulk Majorana mass, diluted by the large hierarchy $\mu_1p$ in the denominator. Other orderings of $p,\mu_1,M_j$ can be worked through by the same method starting from Eq.~\eqref{eq-app:mbb_general}, but are not relevant to this work. A summary of interesting ones (valid as long as $m_{D,j}$ is the smallest scale) is given by
\begin{equation}
    p^2\mathrm{Re}\left[-\frac1{f^{(j)}(ip)}\right] \approx \left(\frac{m_{D,j}^2}{M_j}\right)\times
\begin{cases}
\dfrac{\pi\,  M_j^2}{\mu_1 p}\,, & p \gg M_j,\,\mu_1 \,,\\[10pt]
\dfrac{\pi M_j }{\mu_1}\,, & M_j \gg p,\,\mu_1\,, \\[10pt]
\left(\dfrac{ M_j}{p}\right)^2\,, & \mu_1 \gg p \gg M_j\,, \\[10pt]
1\,, & \mu_1 \gg M_j \gg p\,.
\end{cases}
\end{equation}

\bigskip
Let us explain the origin of this suppression.
Since the active flavour state carries no bare Majorana mass due to gauge invariance, the diagonalizing matrix $V$ obeys $\mathbf M_{00}^{(j)}=\sum_m (V_{0m}^{j})^2\lambda_{m,j}=0$ for each flavour $j$. Writing $\mathcal N_m\equiv V_{0m}$ and omitting the $j$ index, this translates into the exact sum rule
\begin{equation}
\sum_m \mathcal N_m^2\,\lambda_m \;=\;0 .
\label{eq:sumrule}
\end{equation}
We then split the spectrum into two pieces: the lightest seesaw-like eigenstate $\lambda_0$ and the massive KK tower.
We define the KK unperturbed masses by $D_n\equiv\sqrt{M^2+(n\mu_1)^2}$, $n=1,2,\dots$, each contributing a mirror pair $\lambda_{\pm n}\approx\pm D_n$. We solve the two contributions separately.

\medskip
\noindent\textit{Lightest mode.} Solving $f(\lambda)=0$ perturbatively in $m_{D}^2$ near $\lambda=0$,
\begin{equation}
\mathcal N_0^2\,\lambda_0 \;\approx\; \frac{\pi m_{D}^2}{\mu_1}\coth\!\left(\frac{\pi M}{\mu_1}\right),
\label{eq:light}
\end{equation}
which reduces to the ordinary seesaw value $m_D^2/M$ for $M\ll\mu_1$.

\medskip
\noindent\textit{Lightest mode partner.} A second root of $f(\lambda)=0$, of the same order in $m_D^2$, is obtained by perturbing around the unperturbed pole at $\lambda=-M$. We label the next-to-lightest mode as $0^+$. At leading order in the $m_D$ expansion, this gives
\begin{equation}
\mathcal N_{0^+}^2\,\lambda_{0^+} \;\approx\; -\frac{m_{D}^2}{M},
\label{eq:partner}
\end{equation}
independent of $\mu_1$ to this order. It is comparable to Eq.~\eqref{eq:light} and nearly cancels it when $M\ll\mu_1$ (seesaw limit), but is suppressed relative to it by $\mu_1/M$ when $M\gg\mu_1$, where it may be dropped.
\medskip
\noindent\textit{Massive tower.} Non-degenerate perturbation theory around each unperturbed level $D_n$ gives
\begin{equation}
\mathcal N_{+n}^2\lambda_{+n} \approx \frac{m_{D}^2(D_n-M)}{D_n^2},
\qquad
\mathcal N_{-n}^2\lambda_{-n} \approx -\frac{m_{D}^2(D_n+M)}{D_n^2},
\end{equation}
so each pair contributes
\begin{equation}
\mathcal N_{+n}^2\lambda_{+n}+\mathcal N_{-n}^2\lambda_{-n} \;\approx\; -\frac{2m_{D}^2 M}{D_n^2} \;\sim\; \mathcal O(1/n^2)\,.
\label{eq:pair}
\end{equation}
The $1/n^2$ term makes the sum absolutely convergent.
\medskip
\noindent\textit{Truncated partial sums.} We now define, for an arbitrary cutoff level $N$, the partial sums
\begin{align}
&S(N)\equiv \mathcal N_0^2\lambda_0+\mathcal N_{0^+}^2\lambda_{0^+}+\sum_{n=1}^{N}\Big[\mathcal N_{+n}^2\lambda_{+n}+\mathcal N_{-n}^2\lambda_{-n}\Big]\,,
&
T(N)\equiv \sum_{n>N}\Big[\mathcal N_{+n}^2\lambda_{+n}+\mathcal N_{-n}^2\lambda_{-n}\Big]\,.
\end{align}
The function $T(N)$ quantifies the residual of the exact sum rule of Eq.~\eqref{eq:sumrule}, which sets $S(N)+T(N)=0$. Nevertheless, for $N$ sufficiently large, that is also the regime where we can better approximate the eigenvalues of the mass matrix $\lambda_n$.
Using Eq.~\eqref{eq:pair} and $\sum_{n>N}1/n^2\approx 1/N$ for $N\gg M/\mu_1$, we find
\begin{align}
&T(N)\;\approx\;-\frac{2m_D^2 M}{\mu_1^2\,N}\,, &&\Longrightarrow &&S(N)\;\approx\;\frac{2m_D^2 M}{\mu_1^2\,N}\,.
\label{eq:tail}
\end{align}
As $N\to\infty$, both contributions correctly vanish.
Within the context of $0\nu\beta\beta$ the result is quite interesting.
Setting $N$ to a meaningful physical cutoff such as $N=p/\mu_1$ (i.e.\ truncating at $\lambda_n\approx p$) gives $S(p),T(p)=\pm\,2m_D^2M/(\mu_1 p)$, the same order as the total $m_{\beta\beta}\simeq \pi m_D^2M/(\mu_1 p)$ in the $p\gg M,\mu_1$ regime: light and heavy sectors contribute comparably; neither dominates.
\section{Constraints in Hierarchical Regimes}
\label{app:hierarchical}
\begin{figure*}[t]
  \centering
  \begin{subfigure}[t]{0.32\textwidth}
    \centering
    \includegraphics[width=\textwidth]{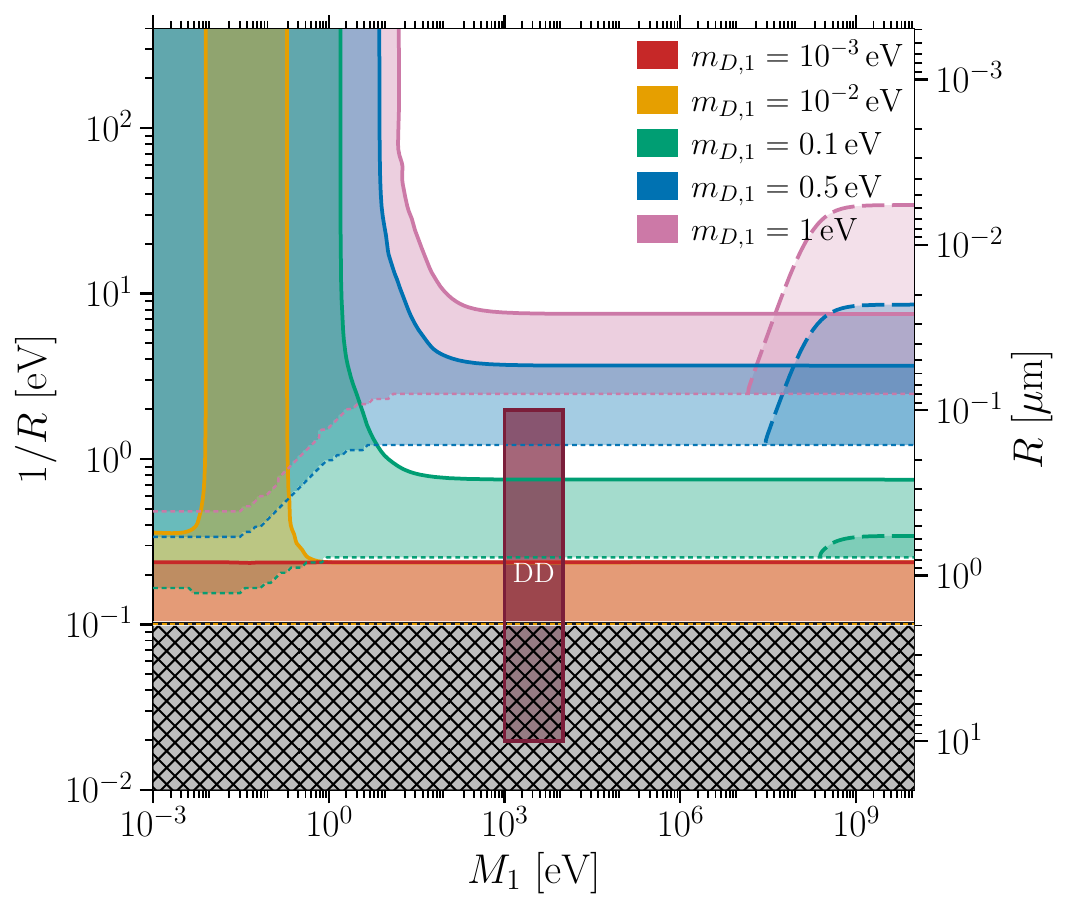}
    \caption{Generation 1.}
    \label{fig:onegen-gen1}
  \end{subfigure}
  \hfill
  \begin{subfigure}[t]{0.32\textwidth}
    \centering
    \includegraphics[width=\textwidth]{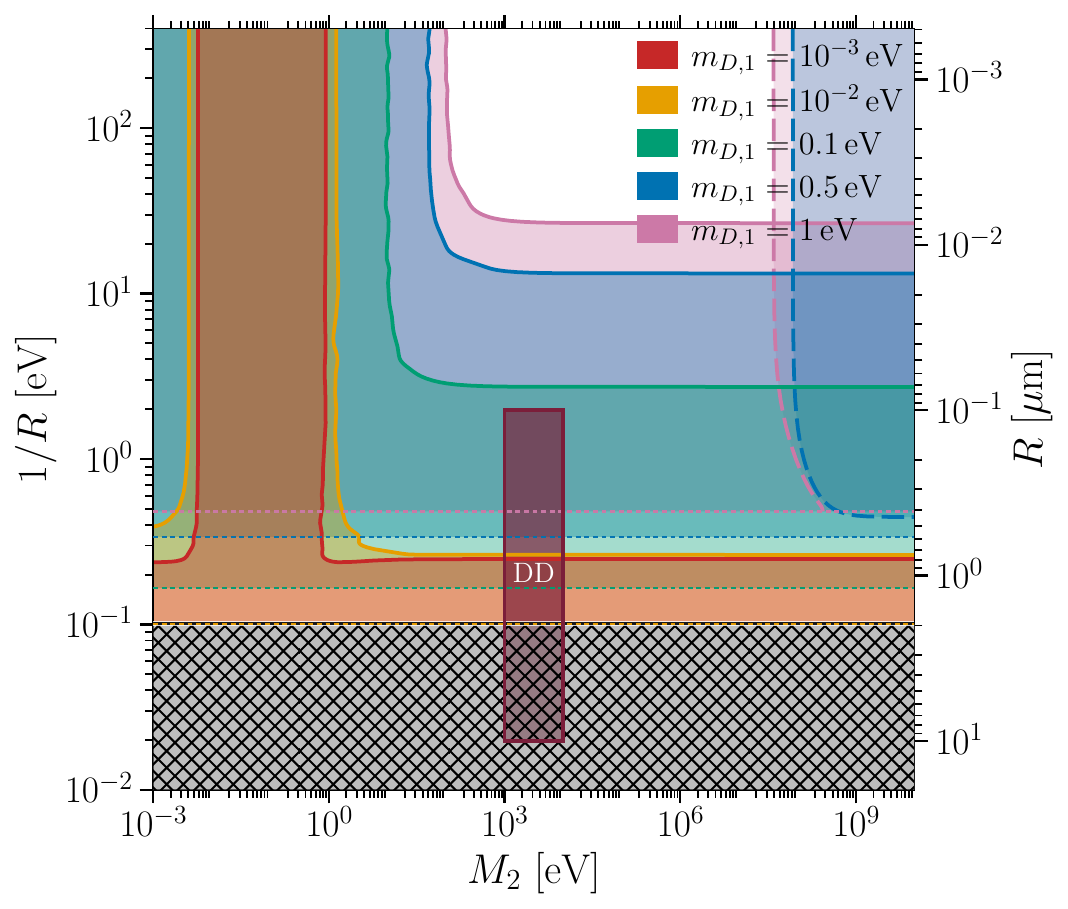}
    \caption{Generation 2.}
    \label{fig:onegen-gen2}
  \end{subfigure}
  \hfill
  \begin{subfigure}[t]{0.32\textwidth}
    \centering
    \includegraphics[width=\textwidth]{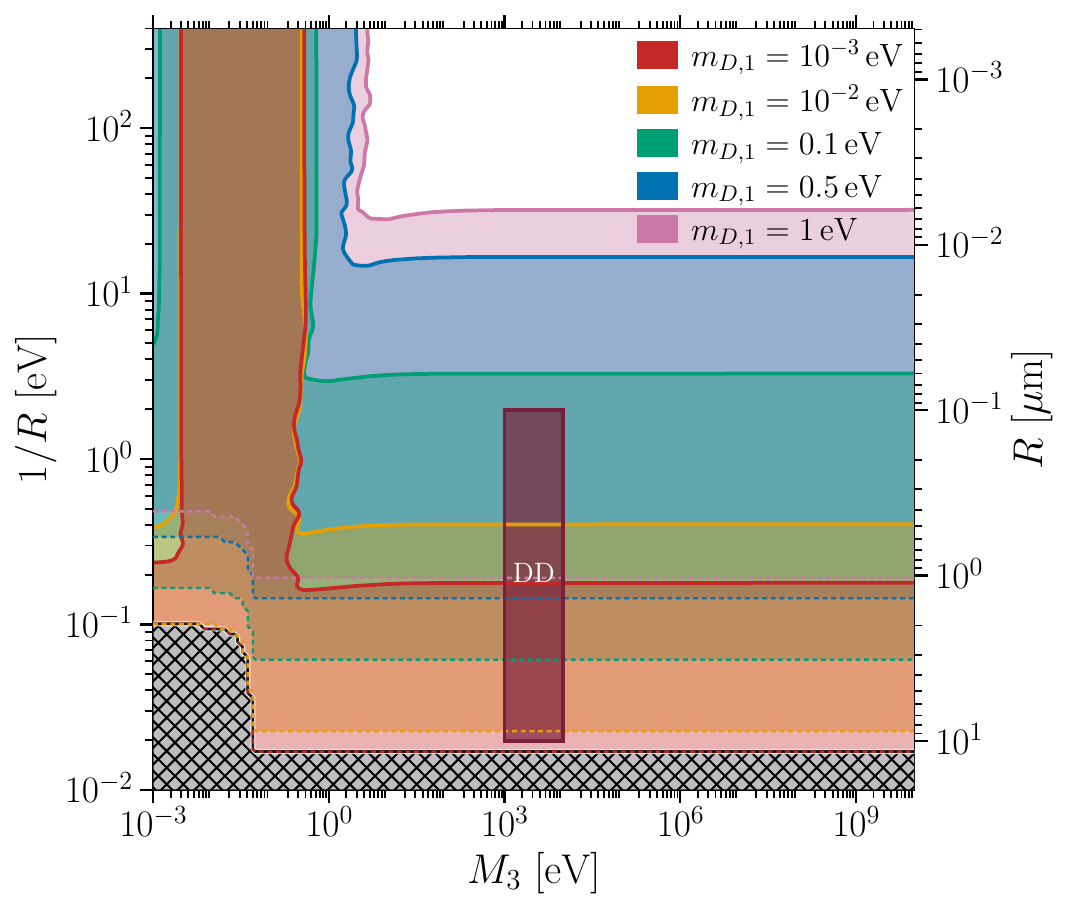}
    \caption{Generation 3.}
    \label{fig:onegen-gen3}
  \end{subfigure}

  \caption{One-generation Majorana-mass at 90\% C.L. limits in NO.   Colours denote $m_{D,1}=10^{-3},10^{-2},0.1,0.5,$ and $1$~eV. Solid contours show the Daya Bay recast and dashed contours the conservative KamLAND-Zen $^{136}$Xe envelope. The cross-hatched region cannot reproduce observed mass splitting for any displayed slice; the coloured thin boundaries mark the additional slice-dependent no-solution regions. The maroon rectangle marks the DD window.}
  \label{fig:onegen-NO}
\end{figure*}

\begin{figure*}[t]
  \centering
  \begin{subfigure}[t]{0.32\textwidth}
    \centering
    \includegraphics[width=\textwidth]{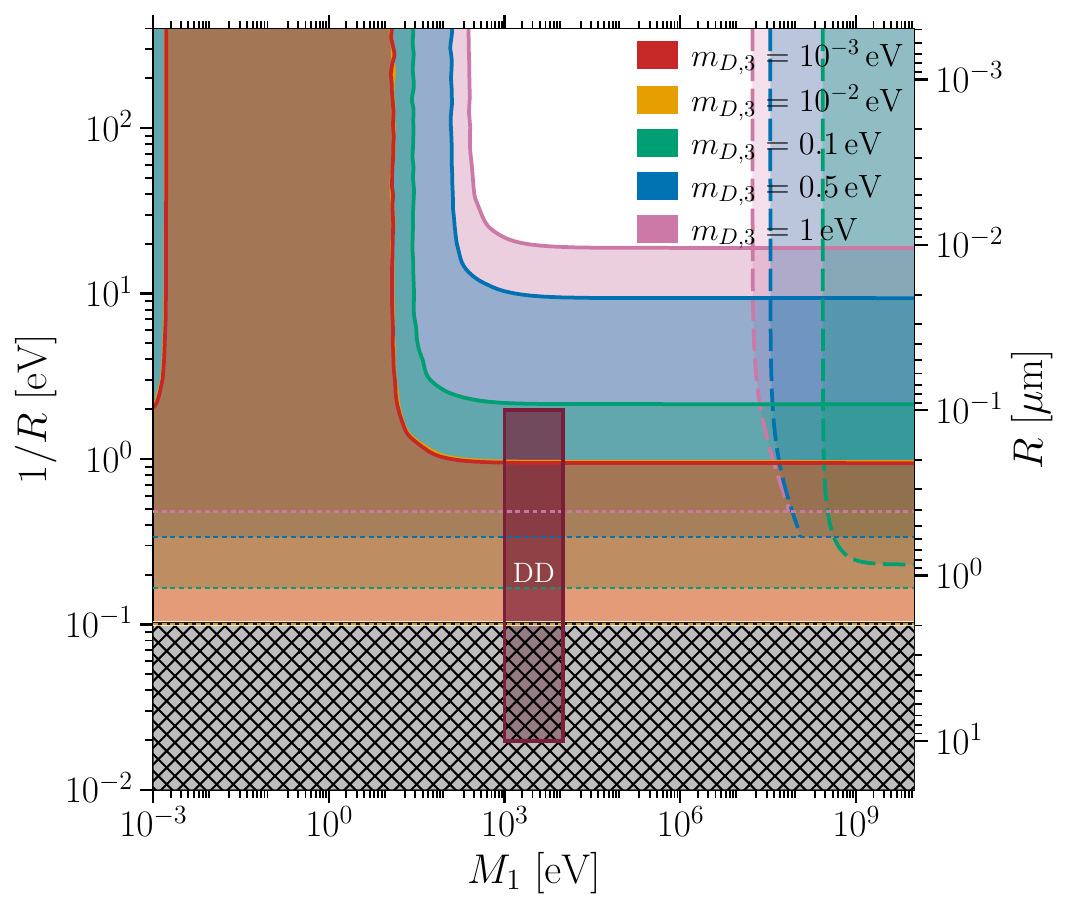}
    \caption{Generation 1.}
    \label{fig:onegen-IO-gen1}
  \end{subfigure}
  \hfill
  \begin{subfigure}[t]{0.32\textwidth}
    \centering
    \includegraphics[width=\textwidth]{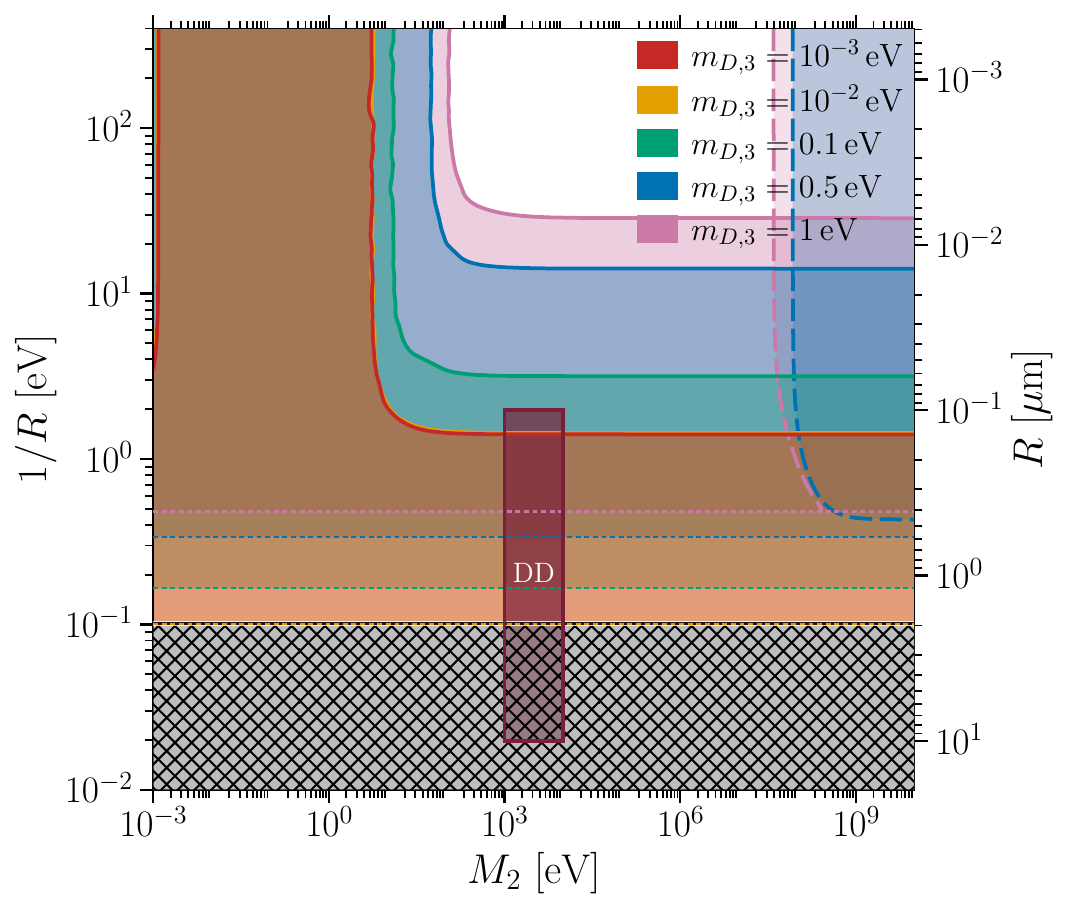}
    \caption{Generation 2.}
    \label{fig:onegen-IO-gen2}
  \end{subfigure}
  \hfill
  \begin{subfigure}[t]{0.32\textwidth}
    \centering
    \includegraphics[width=\textwidth]{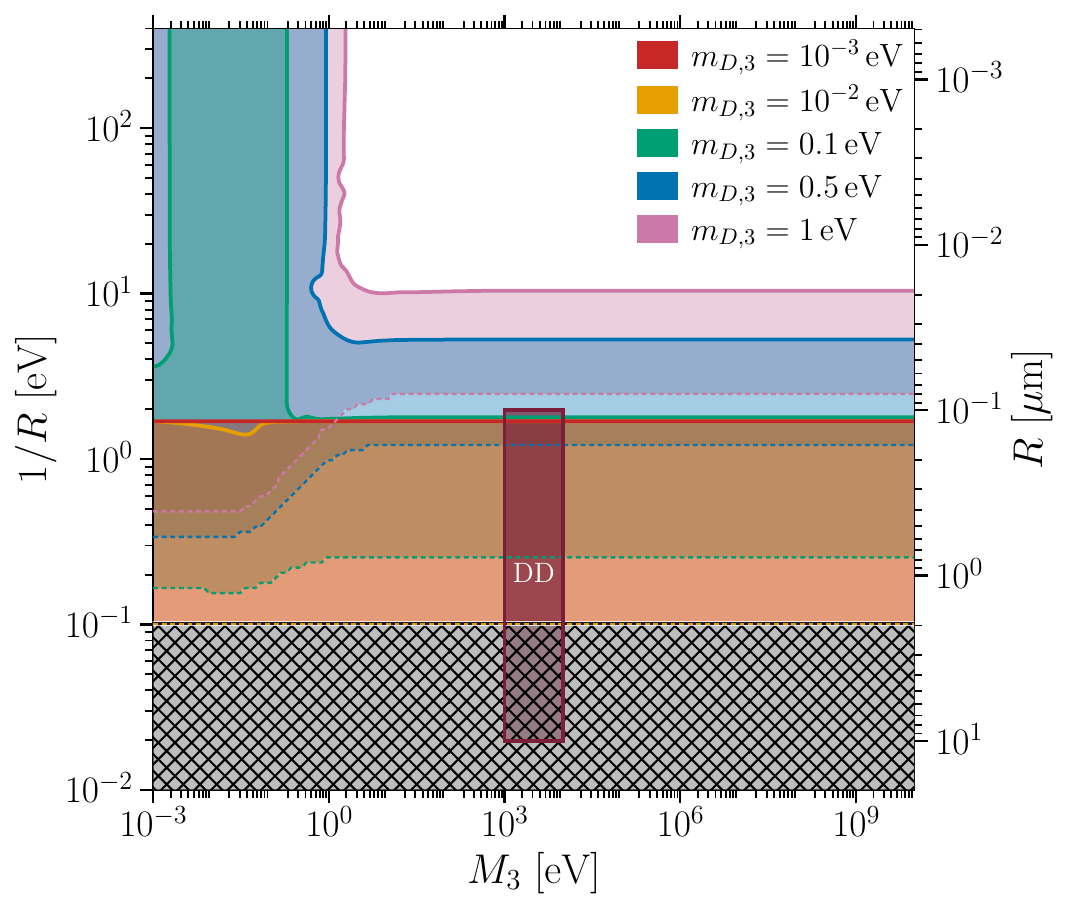}
    \caption{Generation 3.}
    \label{fig:onegen-IO-gen3}
  \end{subfigure}
  \caption{As in Fig.~\ref{fig:onegen-NO}, but for IO and $m_D\equiv m_{D,3}$. }
  \label{fig:onegen-IO}
\end{figure*}

In the main text, we present the result for the simplified case study in which all Majorana masses across the three flavours are identical. Here we depart from such an assumption and investigate the impact of hierarchies.
The results are shown in Figs.~\ref{fig:onegen-NO}-\ref{fig:onegen-IO} for NO and IO, respectively. The three panels show on the $x$-axis the generation carrying the appreciable Majorana mass. The other two are fixed to $10^{-9}$~eV as a numerical $0^+$ limit. In the NO panels, the displayed Dirac mass is $m_{D,1}$; in the IO panels (Fig.~\ref{fig:onegen-IO}) it is $m_{D,3}$, consistent with the choice made in Sec.~\ref{sec:results} and in the caption of Fig.~\ref{fig:main-IO}. We derive the other two Dirac masses from the measured splittings rather than scanning them independently. Unlike the other benchmark cases,  KATRIN is not shown because the quasi-Dirac spectator towers place additional thresholds inside its fit window.

\bibliographystyle{BiblioStyle}
\bibliography{Biblio_Draft}
\end{document}